\documentclass[preprint,12pt]{elsarticle}

\usepackage{hyperref}
\usepackage{graphicx}
\usepackage{amsfonts}
\usepackage{amsmath}
\usepackage[toc,page]{appendix}

\journal{Applied Mathematics and Computation}  

\begin{document}

\begin{frontmatter}

\title{A new insight into the consistency of the SPH interpolation formula}


\author[address1]{Leonardo Di G. Sigalotti\corref{mycorrespondingauthor}}
\cortext[mycorrespondingauthor]{Corresponding author}
\ead{leonardo.sigalotti@gmail.com}

\author[address2,address3]{Otto Rend\'on}
\ead{ottorendon@gmail.com}

\author[address4]{Jaime Klapp}
\ead{jaime.klapp@inin.gob.mx}

\author[address1]{Carlos A. Vargas}
\ead{carlovax@gmail.com}

\author[address1]{Fidel Cruz}
\ead{fidelcru@gmail.com}

\address[address1]{Area of Physics of Irreversible Processes, Department of
Basic Sciences, Autonomous Metropolitan University - Azcapotzalco (UAM-A),
Av. San Pablo 180, 02200 Mexico City, Mexico}

\address[address2]{Center of Physics, Venezuelan Institute of Scientific
Research (IVIC), Apartado Postal 20632, Caracas 1020-A, Venezuela}

\address[address3]{Physics Department, Faculty of Science and Technology,
University of Carabobo (UC), Valencia, Carabobo State, Venezuela}

\address[address4]{Physics Department, National Institute of Nuclear Research
(ININ), Carretera Mexico-Toluca km. 36.5, La Marquesa, 52750 Ocoyoacac, State 
of Mexico, Mexico}


\begin{abstract}
In this paper, the consistency of the smoothed particle hydrodynamics (SPH)
interpolation formula is investigated by analytical means. A novel error analysis is
developed in $n$-dimensional space using the Poisson summation formula, which enables the
simultaneous treatment of both the kernel and particle approximation errors for arbitrary
particle distributions. New consistency integral relations are derived for the particle
approximation, which correspond to the cosine Fourier transform of the kernel
consistency conditions. The functional dependence of the error
bounds on the SPH interpolation parameters, namely the smoothing length, $h$, and the number
of particles within the kernel support, ${\mathcal{N}}$, is demonstrated explicitly from which
consistency conditions arise. As ${\mathcal{N}}\to\infty$, the particle
approximation converges to the kernel approximation independently of $h$ provided that
the particle mass scales with $h$ as $m\propto h^{\beta}$ with $\beta >n$, where $n$ is
the spatial dimension. This implies that as $h\to 0$, the joint limit $m\to 0$,
${\mathcal{N}}\to\infty$, and $N\to\infty$ is necessary for complete convergence to the
continuum, where $N$ is the total number of particles. The analysis also reveals
a dominant error term of the form $(\ln {\mathcal{N}})^{n}/{\mathcal{N}}$ for finite
${\mathcal{N}}$, as it has long been conjectured based on the similarity between the SPH
and the quasi-Monte Carlo estimates. When ${\mathcal{N}}\gg 1$, the error of the SPH interpolant
decays as ${\mathcal{N}}^{-1}$ independently of the dimension. This ensures approximate partition 
of unity of the kernel volume.
\end{abstract}


\end{frontmatter}


\section{Introduction}

Smoothed particle hydrodynamics (SPH) is a Lagrangian particle method that was
developed in the late 70s for the simulation of astrophysical flows \cite{Lucy77,
Gingold77}. However, it has emerged in recent years as a promising numerical technique
for the simulation of complex fluid flows as well as for a large variety of problems in
computational mechanics and related areas \cite{Monaghan05,Rosswog09,Liu10,Monaghan12}.
Given the widespread use of SPH today, a complete understanding of the errors
is mandatory to account for the lack of consistency of the standard SPH approximation.
The mathematical concept of consistency is related to how
closely the numerical discrete equations approximate the exact equations. In other
words, consistency is a measure of the local truncation error. In SPH the discrete
equations are assembled by replacing the field functions and gradients in the exact
partial differential equations by their basic SPH interpolation formulae. Therefore, it
is of interest to derive the functional dependence of their local
truncation errors on the interpolation parameters, namely the smoothing length, $h$,
and the number of neighbors within the kernel support, ${\mathcal{N}}$.

Although significant progress has been done over the years to restore SPH consistency
(i.e., exact interpolation of low-order polynomials)
\cite{Bonet99,Chen99,Liu03,Zhang04,Liu06,Litvinov15,Sibilla15}
and investigate the truncation errors carried by the SPH summation interpolants
\cite{Ben00,Quinlan06,Vaughan08,Read10,Fatehi11,Zhu15}, their explicit functional 
dependence is not known and their actual nature is understood only in heuristic 
terms. For example, it is still unclear how the second-order accuracy noted by many
authors for the continuous kernel approximation translates into the full discrete
form, making a difficult task to provide simple general statements about the accuracy
and convergence of the SPH interpolation. On the other hand, the convergence of SPH to
the exact fluid-dynamics equations has been proved in Sobolev norms with respect to
suitable regularizations of the pressure field \cite{DiLisio97} and with respect to the
Wasserstein distance between measures as the number of particles tends to infinity
\cite{DiLisio98,Zisis16}, while it has been proved numerically by using a modified 
transport-velocity formulation \cite{Adami13} and by relaxing the particle distributions 
toward satisfying partition of unity \cite{Litvinov15}. A consistent SPH approach for
the simulation of interfacial multiphase flows has also recently appeared \cite{Krimi18},
which includes a surface tension formulation that reproduces the tangential properties
of the tensor surface stress.

The development of the theory of meshless methods has been motivated by the fact that
mesh-free and mesh-adaptive discretizations are often better suited to cope with
geometric changes of the domain of interest, as for the case of free surfaces and
large deformations, than the classical structured-mesh discretization techniques. 
Several mesh-free methods have been proposed since SPH was born. Among the most
widely used in engineering applications, we find the element-free Galerkin method
(EFG) \cite{Belytschko94}, the reproducing kernel particle method (RKPM) \cite{Liu95}, 
the moving least squares method (MLSM) \cite{Liu97}, and their improved extensions
GEFG, GRKPM \cite{Shodja07,Hashemian08}, and GMLSM \cite{Atluri99}, respectively.
All these methods are based on a set of independent points (or particles) and use
a shape (or kernel) function to approximate a continuous function and so they share
similarities with the SPH method. The EFG method has been successfully used to
solve a great number of problems in solid mechanics, with very few applications to 
fluid flow problems \cite{Singh05,Staroszczyk13}. In particular, the RKPM method has
been used recently to explore the vibration performance of double-walled carbon
nanotubes \cite{Kiani13,Kiani14,Kiani15}. On the other hand, GRKPM simulations of
concrete deterioration due to reinforcement corrosion has shown very good agreement
with experimentally observed data \cite{Shodja10}. Moreover, the dynamical response
of multispan viscoelastic thin and deformable beams under the excitation of a moving 
mass has also been successfully studied using the GMLSM method \cite{Kiani09,Kiani10}. 

In this article we provide a new mathematical analysis to investigate the truncation 
errors carried by the SPH estimate of a function using the Poisson summation formula.
The analysis can also be applied to evaluate the accuracy and convergence of other 
meshless methods as the ones mentioned above, including the quasi-Monte Carlo method. 
The Poisson formula was first used by Monaghan \cite{Monaghan05} in an attempt to 
estimate the errors in the SPH summation interpolant for a linear function in one
dimension using equidistant particles and a Gaussian kernel. However, no conclusions 
were reached about the consistency of the method from his analysis.
In contrast, here we provide expressions for the error bounds of the SPH estimate of an
arbitrary function for non-uniformly distributed sets of particles. These expressions 
account for full consistency of the SPH summation interpolants and give the functional 
dependence of the error bounds on the SPH interpolation parameters. This article is 
organized as follows. Section 2 contains some preliminary definitions, while an overview 
of the SPH interpolation theory is given in Section 3, where key mathematical constructs 
are introduced for use in the error analysis. Section 4 deals with the error analysis of 
the SPH interpolation formula in one-space dimension ($n=1$) and Section 5 extends the 
analysis to $n$-dimensions. Finally, a discussion of the results is given in Section 6 
and the conclusions are summarized in Section 7.

\section{Preliminary definitions}

The Poisson summation formula is valid for all test functions 
$f({\bf x})\in {\mathcal{S}}(\mathbb{R}^{n})$ \cite{Schwartz66}. Here $\mathbb{R}^{n}$ is 
the $n$-dimensional Euclidean space, where the length of vector
${\bf x}=(x_{1},x_{2},\ldots ,x_{n})$ is defined by the Euclidean norm
$\parallel {\bf x}\parallel _{2}:=\sqrt{x_{1}^{2}+x_{2}^{2}+\cdots +x_{n}^{2}}$. Let us
also denote by $\mathbb{N}$ and $\mathbb{Z}$ the fields of all natural and integer numbers,
respectively. The following definitions on the function spaces and compact support of a
function are used.

\vspace{0.4cm}

{\bf Definition 1} (Continuous function spaces). {\it ${\cal{S}}(\mathbb{R}^{n})$ denotes
the Schwartz space of all infinitely continuous functions on $\mathbb{R}^{n}$ with fast
decay at infinity along with all derivatives}.

\vspace{0.4cm}

{\bf Definition 2} (Dual function spaces). {\it ${\cal{S}}^{\prime}(\mathbb{R}^{n})$ is
the dual space of ${\cal{S}}(\mathbb{R}^{n})$, which is also a subspace of
${\cal{D}}^{\prime}(\mathbb{R}^{n})$, i.e., the dual space of ${\cal{D}}(\mathbb{R}^{n})$,
which is the space of all smooth functions with compact support on $\mathbb{R}^{n}$. Every
function of $\cal{D}$ belongs to $\cal{S}$}.

\vspace{0.4cm}

{\bf Definition 3} (Compact support of a function). {\it The support of a function
$f({\bf x})\in {\cal{D}}(\mathbb{R}^{n})$ that is locally integrable in $\mathbb{R}^{n}$
is the closure $\Gamma\in {\rm supp}(f)$ of the set of points ${\bf x}$ such that
$f({\bf x})\neq 0$}.

\section{SPH interpolation theory}

The SPH interpolation involves a two step procedure. The first is known as the
{\it kernel approximation} and the second is known as the {\it particle
approximation} \citep{Monaghan05}.

\subsection{Kernel approximation}

Using ideas from distribution theory, the kernel approximation of a smooth
function, $f({\bf x}) : \mathbb{R}^{n}\to\mathbb{R}$, is built up from the
Dirac-$\delta$ sampling property by approximating the Dirac-$\delta$ distribution
with a continuous kernel function $W$ \cite{Liu10} such that
\begin{equation}
\langle f({\bf x})\rangle =\int _{\Omega _{n}}f({\bf x}^{\prime})
W(\parallel {\bf x}-{\bf x^{\prime}}\parallel ,h)d^{n}{\bf x}^{\prime},
\end{equation}
where $\Omega _{n}\subset {\mathbb{R}}^{n}$ is the spatial domain and $h$ is
the width of the kernel, most commonly known as the smoothing length. The
notation $\langle f({\bf x})\rangle$ is used to denote the kernel estimate of
$f({\bf x})$. The kernel function in relation (1) must fulfill the
normalization condition
\begin{equation}
M_{0}=\int _{\Omega _{n}}W(\parallel {\bf x}-{\bf x}^{\prime}\parallel ,h)
d^{n}{\bf x}^{\prime}=1,
\end{equation}
and must be positive definite, symmetric, monotonically decreasing, and
tend to $\delta ({\bf x}-{\bf x}^{\prime})$ as $h\to 0$ so that
$\langle f({\bf x})\rangle\to f({\bf x})$. Suitable kernels must also have
a compact support so that $W=0$ if
$\parallel {\bf x}-{\bf x}^{\prime}\parallel >kh$, where $k$ is some number
that specifies the support of the kernel.

With the use of Taylor series expansions, many authors have noted that the
kernel approximation (1) has a leading second-order error $O(h^{2})$ when $h$ is
not in the limit. If in the integral approximant (1) we expand $f({\bf x}^{\prime})$
in Taylor series about ${\bf x}$ and use relation (2), the kernel approximation 
becomes
\begin{equation}
\langle f({\bf x})\rangle =f({\bf x})+\sum _{l=1}^{\infty}\frac{1}{l!}
\nabla ^{(l)}f({\bf x})::\cdots :\int _{\Omega _{n}}
({\bf x}^{\prime}-{\bf x})^{l}W(\parallel {\bf x}-
{\bf x}^{\prime}\parallel ,h)d^{n}{\bf x}^{\prime},
\end{equation}
where $\nabla ^{(l)}$ denotes the product of the $\nabla$ operator $l$ times with
respect to coordinates ${\bf x}$, the symbol ``$::\cdots :$'' denotes the $l$th-order
inner product, and $({\bf x}^{\prime}-{\bf x})^{l}$ is a tensor of rank $l$.  From
expansion (3) it follows that exact interpolation of a polynomial of
order $m$ (i.e., consistency $C^{m}$) can be obtained if the family of consistency
relations (or moments of the kernel) \citep{Liu06}
\begin{equation}
{\bf M}_{l}=\int _{\Omega _{n}}({\bf x}^{\prime}-{\bf x})^{l}W(\parallel {\bf x}-
{\bf x}^{\prime}\parallel ,h)d^{n}{\bf x}^{\prime}={\bf 0}^{(l)},
\end{equation}
are exactly fulfilled for $l=1,2,\cdots ,m$, where ${\bf 0}^{(1)}=(0,0,0)$ is the
null vector and ${\bf 0}^{(l)}$ is the zero tensor of rank $l$.
$C^{0}$ consistency of the kernel approximation is always guaranteed because of the
normalization condition (2), while relations (4) are always satisfied for $l=1$ due to
the symmetry of the kernel and therefore $C^{1}$ consistency is also automatically
ensured. The same is true for all odd $l\geq 3$. Only for $l$ even the integrals
(4) contribute with finite sources of error unless
$W(\parallel {\bf x}-{\bf x}^{\prime}\parallel ,h)\to\delta ({\bf x}-{\bf x}^{\prime})$.
The second-order error follows from the $l=2$ non-vanishing term in expansion (3).
Using Eq. (1) it follows that \citep{Sigalotti16}
\begin{equation}
{\bf M}_{2}=\langle {\bf x}{\bf x}\rangle -\langle {\bf x}\rangle\langle {\bf x}\rangle
\neq {\bf 0}^{(2)},
\end{equation}
provided that consistencies $C^{0}$ and $C^{1}$ are achieved. This term is just the
variance of the position of the interpolation points (particles) and is a measure of the
spread in position relative to the mean. Thus $C^{2}$ consistency is not achieved by
the kernel approximation unless
$W(\parallel {\bf x}-{\bf x}^{\prime}\parallel ,h)\to\delta ({\bf x}-{\bf x}^{\prime})$.
The form of the second moment in (5) bears a tight resemblance to the expectation value
derived by Di Lisio, Grenier, and Pulvirenti \cite{DiLisio98} for the SPH convergence of
a sequence of empirical measures when $N\to\infty$. However, note that upon choosing a
kernel function with vanishing ${\bf M}_{2}$, $C^{2}$ consistency can be achieved for the
kernel approximation.

A similar analysis for the kernel estimate of the gradient, namely
\begin{equation}
\langle\nabla f({\bf x})\rangle =\int _{\Omega _{n}}f({\bf x}^{\prime})\nabla
W(\parallel {\bf x}-{\bf x}^{\prime}\parallel ,h)d^{n}{\bf x}^{\prime},
\end{equation}
leads to the Taylor series expansion
\begin{equation}
\langle\nabla f({\bf x})\rangle =\sum _{l=0}^{\infty}\frac{1}{l!}
\nabla ^{(l)}f({\bf x})::\cdots :
\int _{\Omega _{n}}({\bf x}^{\prime}-{\bf x})^{l}\nabla
W(\parallel {\bf x}-{\bf x}^{\prime}\parallel ,h)d^{n}{\bf x}^{\prime},
\end{equation}
where the moments of the kernel gradient must satisfy the following conditions
to achieve $C^{m}$ consistency
\begin{eqnarray}
{\bf M}_{0}^{\prime}&=&\int _{\Omega _{n}}\nabla W(\parallel {\bf x}-{\bf x}^{\prime}
\parallel ,h)d^{n}{\bf x}^{\prime}={\bf 0}^{(1)},\nonumber\\
{\bf M}_{1}^{\prime}&=&\int _{\Omega _{n}}({\bf x}^{\prime}-{\bf x})\nabla
W(\parallel {\bf x}-{\bf x}^{\prime}\parallel ,h)d^{n}{\bf x}^{\prime}={\bf I},\\
{\bf M}_{l}^{\prime}&=&\int _{\Omega _{n}}({\bf x}^{\prime}-{\bf x})^{l}\nabla
W(\parallel {\bf x}-{\bf x}^{\prime}\parallel ,h)d^{n}{\bf x}^{\prime}={\bf 0}^{(l+1)},
\nonumber
\end{eqnarray}
for $l=2,3,\cdots ,m$, where ${\bf I}$ is the unit tensor.

\subsection{Particle approximation}

If the spatial domain $\Omega _{n}$ is divided into $N$ sub-domains, labeled
$\Omega _{a}$, each of which encloses an interpolation point (or particle) $a$ at
position ${\bf x}_{a}\in\Omega _{a}$, the discrete equivalent of Eq. (1) is defined by
\begin{equation}
f_{a}=\sum _{b=1}^{\cal{N}}f_{b}W_{ab}\Delta V_{b},
\end{equation}
where $W_{ab}=W(\parallel {\bf x}_{a}-{\bf x}_{b}\parallel ,h)$,
$\Delta V_{b}$ is the volume of sub-domain $\Omega _{b}$, and the summation is over 
$\cal{N}$ points within the support of the kernel of spherical volume
${\cal{V}}_{n}=B_{n}(kh)^{n}/n$, where $B_{n}=2\pi ^{n/2}/\Gamma (n/2)$ is the solid 
angle in $n$-dimensional Euclidean space subtended by the complete $(n-1)$-dimensional 
spherical surface and $\Gamma$ is the Gamma function. This gives the exact result of 
$4\pi$ steradians for $n=3$. In general, the summation interpolant (9) refers to $\cal{N}$ 
non-uniformly distributed points and therefore the volumes $\Delta V_{b}$ may not be the 
same for all particles. In almost all SPH applications, the particle volume $\Delta V_{b}$ 
is replaced by the ratio $m_{b}/\rho _{b}$, where $m_{b}$ and $\rho _{b}$ are the mass and 
density of particle $b$, respectively. For uniformly distributed particles,
Fulk \citep{Fulk94} derived error bounds for the SPH approximation (9) through the use of
Taylor series expansions and proved the following:

\vspace{0.4cm}

{\bf Lemma 1} (Consistency for the SPH Approximation). {\it Given a function,
$f({\bf x})\in {\cal{S}}(\mathbb{R}^{3})$, and given a kernel interpolation function,
$W$, that is symmetric, positive-definite, normalized, and has compact support, the SPH
approximation (9) is consistent with the identity operator, $If({\bf x})=f({\bf x})$,
under the uniform norm,
\begin{eqnarray}
\parallel Sf-If\parallel _{\infty}&=&\parallel Kf-If+Sf-Kf\parallel _{\infty}\nonumber\\
&\leq&\parallel Kf-If\parallel _{\infty}+\parallel Sf-Kf\parallel _{\infty},
\end{eqnarray}
provided that $\Delta V_{b}$ is equal to $m_{b}/\rho _{b}$, where
$If=f({\bf x}_{a})$ is the value of the exact function at ${\bf x}_{a}$, and
$Kf=\langle f({\bf x}_{a})\rangle$ and $Sf=f_{a}$ denote the kernel and the SPH 
approximations of $f({\bf x})$ {\it at} ${\bf x}_{a}$ defined by Eqs. (1) and
(9), respectively}.

\vspace{0.4cm}

Fulk proved that in the limit of vanishing inter-particle distances
$\parallel Sf-Kf\parallel _{\infty}\to 0$, while
$\parallel Kf-If\parallel _{\infty}\to 0$ as $h\to 0$. Lemma 1 is valid in
any dimension. Similar conclusions follow for the SPH approximation of the gradient
of a function. However, a definition of consistency for the particle
approximation based
solely on the equivalence $\Delta V_{b}\to m_{b}/\rho _{b}$ is incomplete because
in general
the integral conditions (2) and (4) in discrete form are not satisfied exactly,
i.e.,
\begin{eqnarray}
{\bf M}_{0}&=&\sum _{b=1}^{\cal{N}}\frac{m_{b}}{\rho _{b}}W_{ab}\neq 1,\\
{\bf M}_{l}&=&\sum _{b=1}^{\cal{N}}\frac{m_{b}}{\rho _{b}}({\bf x}_{b}-{\bf x}_{a})^{l}
W_{ab}\neq {\bf 0}^{(l)},\hspace{0.2cm}{\rm for}\hspace{0.2cm}l=1,2,\ldots ,m.
\end{eqnarray}
The same is true for the discrete form of the integral relations (8), leading to
complete loss of consistency due to the particle approximation. Considering the
analogy between quasi-Monte Carlo and SPH particle estimates, Monaghan
\citep{Monaghan85} first conjectured that for low-discrepancy (i.e., quasi-regular
or quasi-random) sequences of particles, as is indeed the case in SPH simulations,
the error carried by the particle approximation is $O((\ln {\cal{N}})^{n}/\cal{N})$.

The complexity of error behavior in SPH has been highlighted by
Quinlan et al. \cite{Quinlan06} and Vaughan et al. \cite{Vaughan08}. The former
authors used the second Euler-MacLaurin formula to estimate this error for
one-dimensional regularly and irregularly distributed particles. They found that
for regular distributions as $h\to 0$, while maintaining constant the ratio of
particle spacing to smoothing length, $\Delta x/h$, the error decays as $h^{2}$
until a limiting discretization error is reached, which is independent of $h$. If
$\Delta x\to 0$ while maintaining $h$ constant, the error decays at a rate which
depends on the kernel smoothness. When particles are distributed non-uniformly,
decreasing $h$ with constant $\Delta x/h$ results in discretization-limited errors
at best. On the other hand, 
Vaughan et al. \cite{Vaughan08} showed that if $C^{0}$ consistency is not
achieved the error is $O(f(\cal{N}))$, which does not converge with $h$, whereas
if $C^{1}$ consistency is achieved the error goes as $O(h^{2}f(\cal{N}))$. They
concluded that if $f({\cal{N}})\sim (\ln {\cal{N}})^{n}/\cal{N}$, an analytical
solution for the functional dependence of the total number of particles $N$ on $h$
cannot be obtained. However, recently Zhu et al. \cite{Zhu15} derived the
parameterizations $h\propto N^{-1/\beta}$ and ${\cal{N}}\propto N^{1-3/\beta}$
for $\beta\in [5,7]$ based on a balance between the kernel and the particle
approximation errors. For $\beta =6$, this gives $h\propto N^{-1/6}$ and
${\cal{N}}\propto N^{1/2}$. They stated that these scaling relations comply with the 
joint limit $N\to\infty$, $h\to 0$, ${\cal{N}}\to\infty$ as a necessary condition
to achieve full particle consistency \cite{Rasio00}. However, the systematic
increase of the number of neighbors ${\cal N}$ with the total number of particles
$N$ demands changing the interpolation kernel to a compactly supported Wendland-type
function \cite{Wendland95}, which, unlike traditional kernels, is free from the so-called
pairing instability when working with large numbers of neighbors \cite{Dehnen12}.

\section{SPH errors in one-space dimension}

For simplicity, first consider the analysis for a set of irregularly distributed 
particles on the real line. Let $\phi (x)\in {{\cal S}}(\mathbb{R})$ be a test 
function and ${\hat\phi}(j)=\int _{\mathbb R}\phi (x)\exp (-i2\pi jx)dx$ its Fourier
transform, where ${\hat\phi}$ also belongs to ${{\cal S}}(\mathbb{R})$. The 
distributional relation
\begin{equation}
\sum _{b=-\infty}^{\infty}\phi (b)=\sum _{j=-\infty}^{\infty}{\hat\phi}(j)
=\int _{\mathbb{R}}\phi (b)db+2\sum _{j=1}^{\infty}\int _{\mathbb{R}}
\phi (b)\cos (2\pi jb)db,
\end{equation}
defines the Poisson summation formula \cite{Meyer16}, where in the leftmost sum
$b\in\mathbb{Z}$, while in the integrals on the right side $b\in\mathbb{R}$. Here, the
integer $b$ in the leftmost sum belongs to the space of particle labels and the integer
$j$ belongs to the dual of the space of labels.
Setting $\phi (b)=f_{b}W_{ab}\Delta x_{b}$, the leftmost summation becomes
\begin{equation}
\sum _{b=-\infty}^{\infty}\phi (b)\to f_{a}=
\sum _{b=1}^{\cal{N}}\frac{m_{b}}{\rho _{b}}f_{b}W_{ab},
\end{equation}
for any ${\cal{N}}\in\mathbb{N}$. The sum on the
left side of relation (14) is over $\mathbb{Z}$, while the one on the right
side runs over the set $[1, {\cal{N}}]$, which is a subset of both
$\mathbb{N}$ and $\mathbb{Z}$. Since the kernel $W_{ab}$ has compact support
centered at the position of particle $a$, only the ${\cal{N}}$ points within the 
support of $W_{ab}$
will actually contribute to the sum on the right side of Eq. (14). Now setting 
$\phi (b)=(m_{b}/\rho _{b})f(x_{b})W(|x_{a}-x_{b}|,h)$ in the integrals on the
right side of Eq. (13), the Poisson summation formula becomes
\begin{eqnarray}
f_{a}&=&\int _{\Omega _{1}}f(x_{b})W(|x_{a}-x_{b}|,h)\frac{m_{b}}{\rho _{b}}db
\nonumber\\
&+&2\sum _{j=1}^{\infty}\int _{\Omega _{1}}f(x_{b})W(|x_{a}-x_{b}|,h)
\cos (2\pi jb)\frac{m_{b}}{\rho _{b}}db,
\end{eqnarray}
where $\Omega _{1}\in {\rm supp}(W)=[x_{a}-kh,x_{a}+kh]$.
The first integral on the right side of Eq. (15) is the kernel approximation of $f(x)$
at point $x=x_{a}$ provided that the equivalence holds
\begin{equation}
dx_{b}=\frac{m_{b}}{\rho _{b}}db,
\end{equation}
which relates the position of a particle to its label. Integration of relation (16)
over the interval $[x_{a}-kh,x_{a}+kh]$ yields
\begin{equation}
b(x_{a}+kh)-b(x_{a}-kh)=\int _{x_{a}-kh}^{x_{a}+kh}\frac{\rho (x)}{m(x)}dx,
\end{equation}
where $m_{b}=m(x_{b})$ and $\rho _{b}=\rho (x_{b})$. Note that the above relation
stands for non-uniformly spaced particles. For a set of equidistant points with
spacing $\Delta$, Eq. (17) reduces to $x_{b}=b\Delta$ and the Poisson formula for
a uniform distribution is recovered. Since there is a one-to-one correspondence
between the particle position $x_{b}$ and its label $b$, the function $x_{b}=x(b)$ 
is bijective.

Expanding $f(x_{b})$ in Taylor series about $x_{a}$ and
inserting the result in Eq. (15) yields the difference ${\cal{E}}_{S}$ between the value
of the exact function $f(x_{a})$ and its particle approximation $f_{a}$, i.e.,
\begin{eqnarray}
{\cal{E}}_{S}&=&f_{a}-f(x_{a})=
\sum _{l=1}^{\infty}\frac{f^{(l)}(x_{a})}{l!}\int _{\Omega _{1}}
(x_{b}-x_{a})^{l}W(|x_{a}-x_{b}|,h)dx_{b}\nonumber\\
&+&2\sum _{j=1}^{\infty}\sum _{l=0}^{\infty}\frac{f^{(l)}(x_{a})}{l!}
\int _{\Omega _{1}}(x_{b}-x_{a})^{l}W(|x_{a}-x_{b}|,h)\cos (2\pi jb)dx_{b},
\nonumber\\
\end{eqnarray}
where $f^{(l)}(x_{a})$ is the $l$th derivative of $f(x)$ evaluated at $x_{a}$.
According to expansion (3), the first sum on the right side of Eq. (18) is the 
difference ${\cal{E}}_{K}=\langle f(x_{a})\rangle -f(x_{a})$ between the exact 
function and its kernel estimate, while the second sum is the deviation of the 
particle approximation from the kernel estimate
${\cal{E}}_{SK}=f_{a}-\langle f(x_{a})\rangle$. The first new result from
inspection of Eq. (18) is that {\it the particle approximation contributes with error
terms that are proportional to the cosine Fourier transform of the integral
consistency relations} (4). Since the kernel is a symmetric function, only those
terms with $l$ even will contribute to the error. Relation (18) is all we need to 
establish the correct consistency constraints for both the kernel and the particle 
approximations.

The number of neighbors of particle $a$ within the kernel support can
be defined by the floor function
\begin{equation}
{\cal{N}}(x_{a},h)=[b(x_{a}+kh)-b(x_{a}-kh)]+\eta,
\end{equation}
where $\eta =1$ if $x_{a}$ is an interpolation point and $\eta =0$ otherwise.
From Eq. (17), the above definition is equivalent to
\begin{equation}
{\cal{N}}(x_{a},h)=\int _{x_{a}-kh}^{x_{a}+kh}\frac{\rho (x)}{m(x)}dx.
\end{equation}
It is easy to show that for small $h$,
${\cal{N}}(x_{a},h)=2(db/dx_{a})kh+O(h^{3})$. Using relation (17), this implies that
\begin{equation}
{\cal{N}}(x_{a},h)=2\frac{\rho (x_{a})}{m(x_{a})}kh+O(h^{3}).
\end{equation}
Since a necessary condition to achieve full particle
consistency is that ${\cal{N}}(x_{a},h)\to\infty$ as $h\to 0$ and $N\to\infty$
\citep{Zhu15}, satisfaction of this joint limit demands that
$\rho (x_{a})/m(x_{a})\sim h^{-\beta}$ (with $\beta >1$) in Eq. (21). In the continuous
limit the density is an intensive physical variable and therefore the above scaling 
must translate into the requirement that the
particle mass scales with $h$ as $h^{\beta}$ (with $\beta >1$) in order to
ensure that ${\cal{N}}(x_{a},h)\to\infty$ as $h\to 0$ in Eq. (21). This implies the
additional important limit $m\to 0$ as $h\to 0$ as a further condition for consistency
of the particle approximation. From the above scaling for the particle mass it
follows that ${\cal{N}}\propto h^{1-\beta}$, which is the one-dimensional
equivalent of the scaling ${\cal{N}}\propto h^{3-\beta}$ (with $\beta >3$)
suggested by Zhu et al. \citep{Zhu15} in three dimensions.

The double summation in Eq. (18) represents the discretization
errors implied by the particle approximation and is a measure of its deviation 
from the kernel estimate:
${\cal{E}}_{SK}=f_{a}-\langle f(x_{a})\rangle$. For any infinitely
differentiable function $f(x)\in {\cal{S}}(\mathbb{R})$, the limit
${\cal{E}}_{SK}\to 0$ is achieved only if
\begin{eqnarray}
M_{l}^{F}&=&\sum _{j=1}^{\infty}\int _{\Omega _{1}}(x_{b}-x_{a})^{l}
W(|x_{a}-x_{b}|,h)\cos (2\pi jb)dx_{b}\nonumber\\
&=&\lim _{{\cal{N}}\to\infty}\sum _{j=1}^{\cal{N}}
\int _{\Omega _{1}}(x_{b}-x_{a})^{l}W(|x_{a}-x_{b}|,h)\cos (2\pi jb)dx_{b}
\nonumber\\
&=&0,
\end{eqnarray}
$\forall l$ even with $l\geq 0$. These relations represent particle
consistency conditions. Note that in the intermediate equality of Eq. (22)
we have used the regularization criterion (B.11) in Appendix B. In actual 
simulations the contribution of these integrals can be neglected only if 
$\cos (2\pi jb)$ oscillates very rapidly within
${\rm supp}(W)=[x_{a}-kh,x_{a}+kh]$, i.e., when
\begin{equation}
[b(x_{a}+kh)-b(x_{a}-kh)]+\eta ={\cal{N}}(x_{a},h)>\frac{1}{j},
\end{equation}
which implies ${\cal{N}}(x_{a},h)>1$ for $j=1$. Note that if inequality (23)
holds for $j=1$, it will also hold for any $j\geq 2$.

\subsection{Error bounds}

The error of the kernel approximation of $f(x)$ at the position of particle
$a$ is given by the first summation in Eq. (18). Bounds on this
error have been previously derived by Fulk \cite{Fulk94}. However, a derivation
is repeated in Appendix A under the uniform norm
\begin{equation}
\parallel Kf-If\parallel _{\infty}=\parallel {\cal{E}}_{K}\parallel _{\infty}
=\sup _{x_{b}\in\Omega _{1}}|{\cal{E}}_{K}|,
\end{equation}
by retaining only second-order terms in the summation.
The result is
\begin{equation}
|{\cal{E}}_{K}|\leq e_{r}^{(2)}h^{2},
\end{equation}
which implies second-order accuracy for the kernel approximation.
A higher order error is also possible if a kernel that has higher
order vanishing even moments is used.

The second summation in Eq. (18) gives the error of the
particle relative to the kernel approximation. As for the
kernel approximation, bounds on this error are also derived under the uniform
norm
\begin{equation}
\parallel Sf-Kf\parallel _{\infty}=\parallel {\cal{E}}_{KS}\parallel _{\infty}
=\sup _{x_{b}\in\Omega _{1}}|{\cal{E}}_{KS}|.
\end{equation}
The result of this analysis is
\begin{equation}
|{\cal{E}}_{KS}|\leq\frac{4}{\pi}a_{0}k\sum _{l=0}^{\infty}h^{l}{\tilde e}_{r}^{(l)}
\left(\lim _{{\cal{N}}\to\infty}\frac{1}{\cal{N}}\sum _{j=1}^{\cal{N}}
\frac{1}{j}\right).
\end{equation}
where $a_{0}>0$ is an upper bound for the kernel function and 
${\tilde e}_{r}^{(l)}$ is defined in Appendix B, where the intermediate
steps leading to inequality (27) are described. A bound for the term between
parenthesis can be obtained using the following theorem for the estimation of
the Euler-Mascheroni constant $\gamma =0.5572...$ \cite{Young91}:

{\bf Theorem 1} {\it For every natural number} ${\cal{N}}$,
\begin{equation}
\frac{1}{2({\cal{N}}+1)}<\sum _{j=1}^{\cal{N}}\frac{1}{j}-\ln{\cal{N}}-\gamma
<\frac{1}{2{\cal{N}}}.
\end{equation}
The proof of this theorem is given by Young \cite{Young91}. The upper bound is the first
term of an asymptotic expansion which can be used to compute $\gamma$. Solving for the
summation term (i.e., the ${\cal{N}}$th harmonic number) in (28), dividing by ${\cal{N}}$,
and applying the limit when ${\cal{N}}\to\infty$ gives
\begin{equation}
\lim _{{\cal{N}}\to\infty}\frac{1}{\cal{N}}
\sum _{j=1}^{\cal{N}}\frac{1}{j}
\leq\lim _{{\cal{N}}\to\infty}\left[\frac{\gamma}{\cal{N}}+
\frac{\ln {\cal{N}}}{{\cal{N}}}+\frac{1}{2{\cal{N}}^{2}}
\right]=0,
\end{equation}
so that $|{\cal{E}}_{KS}|\to 0$ when ${\cal{N}}\to\infty$ and the particle approximation
converges to the kernel approximation. We note that the logarithmic term on the right 
side of the above inequality provides the dominant error for finite
${\cal{N}}$. This term is just the one-dimensional equivalent of the theoretical upper 
bound of the quasi-Monte Carlo method for low-discrepancy (quasi-random) sets of points. 
Since the limit when ${\cal{N}}\to\infty$ of $\ln{\cal{N}}/{\cal{N}}$ is equal to the 
limit when ${\cal{N}}\to\infty$ of $1/{\cal{N}}$, we have the asymptotic expansion
\begin{equation}
\lim _{{\cal{N}}\to\infty}\frac{1}{{\cal{N}}}\sum _{j=1}^{\cal{N}}
\frac{1}{j}\leq\frac{(1+\gamma)}{{\cal{N}}}
+O\left(\frac{1}{{\cal{N}}^{2}}\right),
\end{equation}
for ${\cal{N}}\to\infty$. Using this result into Eq. (27) and retaining terms up to 
$l=2$ in the expansion yields
\begin{equation}
|{\cal{E}}_{KS}|\leq\frac{4(1+\gamma)a_{0}k}{\pi {\cal{N}}}
\left({\tilde e}_{r}^{(0)}+h{\tilde e}_{r}^{(1)}+h^{2}{\tilde e}_{r}^{(2)}\right).
\end{equation}
This shows that in the limit ${\cal{N}}\to\infty$, the particle discretization error 
vanishes (${\cal{E}}_{KS}\to 0$) and so $f_{a}\to\langle f(x_{a})\rangle$, i.e., the 
particle estimate of the function approaches the kernel estimate independently of $h$.
In inequality (31) the leading term is $\propto 1/{\cal{N}}$, which gives a zeroth-order 
convergence rate even though $h\to 0$.

From inequalities (10), (25), and (31) it follows that the error bound for the full SPH
approximation under the uniform norm is
\begin{eqnarray}
\parallel Sf-If\parallel _{\infty}&\leq&|{{\cal{E}}_{K}}|+|{{\cal{E}}_{KS}}|
\nonumber\\
&\leq&\frac{4(1+\gamma)a_{0}k}{\pi {\cal{N}}}
\left({\tilde e}_{r}^{(0)}+h{\tilde e}_{r}^{(1)}+h^{2}{\tilde e}_{r}^{(2)}\right)
+h^{2}e_{r}^{(2)},
\end{eqnarray}
which expresses the important result that {\it complete consistency for the SPH 
estimate of a function can be guaranteed only when ${\cal{N}}\to\infty$ and $h\to 0$
provided that $N\to\infty$ and $m\to 0$}. As a further remark, note that
the scalings $m\propto h^{\beta}$ and ${\cal{N}}\propto h^{1-\beta}$ imply that
$m\propto {\cal{N}}^{\beta /(1-\beta)}$. Since $\beta >1$, this means that by
increasing ${\cal{N}}$ mass resolution is also improved. As an exercise, in
Appendix C we apply the present method to Monaghan's \cite{Monaghan05} 
one-dimensional SPH convergence analysis for a linear function defined over an
infinite set of equidistant particles. 

\section{SPH errors in $n$-dimensional space}

Let $\Lambda\subset\mathbb{R}^{n}$ be a crystalline lattice and
$\Phi ({\bf x}) : \mathbb{R}^{n}\to\mathbb{R}$ a smooth function of locally finite
support $\Gamma$ belonging to ${{\cal{D}}(\mathbb{R}^{n}})$. The distributional Fourier
transform of $\Phi$, namely $\hat\Phi$, in the dual lattice $\Lambda ^{\star}$ is given
by the $n$-dimensional Poisson's formula \citep{Meyer16}
\begin{equation}
\sum _{b_{1},b_{2},\ldots ,b_{n}\in\Lambda}\Phi (b_{1},b_{2},\ldots ,b_{n})=
\sum _{{\bf j}\in\Lambda ^{\star}}\hat\Phi ({\bf j}),
\end{equation}
where the $n$-plet of integers $(b_{1},b_{2},\ldots ,b_{n})$, with $b_{i}\in\mathbb{Z}$
($i=1,2,\ldots ,n$), denotes the projections of the lattice node (or particle) labels
$b\in\mathbb{Z}^{n}$ on the axes of an $n$-dimensional Cartesian coordinate system
and ${\bf j}=(j_{1},j_{2},\ldots ,j_{n})$.
Setting the summation on the left side of Eq. (33) equal to the summation on the right of 
Eq. (9) for the SPH approximation of a function
$f({\bf x}) : \mathbb{R}^{n}\to\mathbb{R}$ at particle position ${\bf x}_{a}$, Poisson's 
formula becomes
\begin{equation}
f_{a}=\sum _{{\bf j}\in\Lambda ^{\star}}\int _{\Omega _{n}}f({\bf x}_{b})
W(\parallel {\bf x}_{a}-{\bf x}_{b}\parallel ,h)\exp (-i2\pi {\bf j}\cdot {\bf b})
\frac{m_{b}}{\rho _{b}}d^{n}{\bf b},
\end{equation}
where now ${\bf b}\in\mathbb{R}^{n}$ and $\Omega _{n}\in {\rm supp}(W)$ is the integration
domain in $n$-dimensional Euclidean space. Since ${\bf b}={\bf b}({\bf x}_{b})$ is
bijective, it admits the inverse ${\bf x}_{b}={\bf x}_{b}({\bf b})$, which in differential
form becomes $d^{n}{\bf x}_{b}=|{\bf J}_{{\bf x}_{b}}({\bf b})|d^{n}{\bf b}$, where
${\bf J}_{{\bf x}_{b}}({\bf b})$ is the Jacobian matrix of the transformation and
\begin{equation}
\left|{\bf J}_{{\bf x}_{b}}({\bf b})\right|=
\left|\frac{\partial (x_{b_{1}},x_{b_{2}},\ldots ,x_{b_{n}})}
{\partial (b_{1},b_{2},\ldots ,b_{n})}\right|=\frac{m_{b}}{\rho _{b}},
\end{equation}
is its determinant. This is the generalization of the differential form (16) in multiple 
dimensions.

Expanding $f({\bf x}_{b})$ in Taylor series about ${\bf x}_{a}$ and
inserting the result in Eq. (34) yields the error between the particle approximation
$f_{a}$ and the exact value of the function at the position of particle $a$
\begin{eqnarray}
{\cal{E}}_{S}^{(n)}&=&f_{a}-f({\bf x}_{a})\nonumber\\
&=&\sum _{l=1}^{\infty}\frac{1}{l!}\nabla ^{(l)}f({\bf x}_{a})::\cdots :
\int _{\Omega _{n}}({\bf x}_{b}-{\bf x}_{a})^{l}W(\parallel {\bf x}_{a}-{\bf x}_{b}\parallel ,h)
d^{n}{\bf x}_{b}\nonumber\\
&+&\sum _{\substack{{\bf j}\in\Lambda ^{\star}\\{\bf j}\neq {\bf 0}}}^{\infty}
\sum _{l=0}^{\infty}\frac{1}{l!}\nabla ^{(l)}f({\bf x}_{a})::\cdots :
{\bf M}_{l}^{F}({\bf j}),
\end{eqnarray}
where ${\bf 0}$ is the $n$-dimensional null vector and
\begin{equation}
{\bf M}_{l}^{F}({\bf j})=\int _{\Omega _{n}}({\bf x}_{b}-{\bf x}_{a})^{l}
W(\parallel {\bf x}_{a}-{\bf x}_{b}\parallel ,h)\exp (-i2\pi {\bf j}\cdot {\bf b})
d^{n}{\bf x}_{b}.
\end{equation}
The error carried by the particle approximation vanishes provided that
${\bf M}_{l}^{F}({\bf j})={\bf 0}^{(l)}$ for all values of $l$, which is the
generalization in $n$-dimensions of the particle consistency relations (22).

The number of neighbors of particle $a$ within the spherical support of the
kernel is therefore defined by
\begin{equation}
{\cal{N}}({\bf x}_{a},h)=\int _{\Omega _{n}}\frac{\rho ({\bf x})}
{m({\bf x},h)}d^{n}{\bf x},
\end{equation}
which for $h\ll 1$ becomes
\begin{equation}
{\cal{N}}({\bf x}_{a},h)=\frac{B_{n}}{n}\frac{\rho ({\bf x}_{a})}
{m({\bf x}_{a},h)}k^{n}h^{n}+O(h^{n+2}).
\end{equation}
Note that for $n=1$, $B_{n}=B_{1}=2$ and the asymptotic form (39) reduces to the
one-dimensional expression (21). From Eq. (39) it follows that the limit
${\cal{N}}\to\infty$ as $h\to 0$ is satisfied only if the particle mass scales
with $h$ as $h^{\beta}$, with $\beta >n$. This reproduces the scaling
${\cal{N}}\propto h^{3-\beta}$ for $n=3$ as was suggested by Zhu et al. \cite{Zhu15}.
Therefore, in $n$-dimensional space, the scaling relations
$m\propto h^{\beta}$ and ${\cal{N}}\propto h^{n-\beta}$ are necessary conditions
to guarantee complete particle consistency in the limit $h\to 0$. There is a subtle
point behind this scaling: as the volume of the kernel support collapses to a point
with no size at all when $h\to 0$, the mass within the support must also tends to 
zero in the limit to yield a finite
density at that point. In this limit ${\cal{N}}/N\to {\cal V}_{n}/V$
as $N\to\infty$, where $V$ is the finite volume of the system. Since
$N\to V{\cal{N}}/{\cal{V}}_{n}\sim h^{-\beta}$, $N\to\infty$ faster than ${\cal{N}}$
as $h\to 0$, i.e., ${\cal{N}}/N\to 0$ in the transition from the discrete to the
continuous space. This last limit was first noted by Rasio \cite{Rasio00} through
a simple linear analysis of sound wave propagation in one dimension. The above
scalings have implications on the minimum resolvable mass, $M_{\min}={\cal{N}}m$,
i.e., the mass within the kernel support. Since $m\propto h^{\beta}$ and
${\cal{N}}\propto h^{n-\beta}$, this implies that $M_{\min}\propto {\cal{N}}^{n/(n-\beta)}$.
In three-space dimensions ($n=3$), $\beta$ varies between $\beta =5$ for quasi-ordered
particle distributions and $\beta =7$ for random distributions \cite{Zhu15}. With the
intermediate choice of $\beta =6$, the minimum resolvable mass scales with
${\cal{N}}$ as $M_{\min}\propto {\cal{N}}^{-1}$, implying that as ${\cal{N}}$ is
increased mass resolution is effectively improved.

A further parameter that characterizes the SPH interpolation procedure is the
distance between pairs of particles
$\Delta ({\bf x}_{i},{\bf x}_{j})$, which provides a measure of their actual distribution
within the support of the kernel. If there exist
${\cal{N}}({\bf x},h)$ particles within $\Omega _{n}\in {\rm supp}(W)$, then there
will be ${\cal{N}}({\bf x},h)[{\cal{N}}({\bf x},h)-1]/2$ different distances
$\Delta ({\bf x}_{i},{\bf x}_{j})$ between particle pairs, which for an irregularly
distributed set will be bounded as
\begin{equation}
\Delta _{\rm min}\leq\Delta ({\bf x}_{i},{\bf x}_{j})\leq\Delta _{\rm max},
\end{equation}
where $\Delta _{\rm min}$ and $\Delta _{\rm max}$ are, respectively, the minimum
and maximum distances. The mean distance $\Delta _{\rm m}$ is given by
\begin{equation}
\Delta _{\rm m}=\left[\frac{{\cal{V}}_{n}}{{\cal{N}}({\bf x},h)}\right]^{1/n},
\end{equation}
where ${\cal{V}}_{n}=B_{n}(kh)^{n}/n$ is the volume of the kernel support.

\subsection{Error Bounds}

Error bounds for the SPH interpolation in $n$-dimensional space can be
determined under the uniform norm (10) in terms of the sum of
the difference between the kernel approximation of a function and its exact value,
${\cal{E}}_{K}^{(n)}$, given by the first summation in Eq. (36),
and the difference between the kernel and the particle approximations,
${\cal{E}}_{KS}^{(n)}$, represented by the second summation.

The bound of ${\cal{E}}_{K}^{(n)}$ is derived in Appendix D for completeness
and the result is given by the inequality
\begin{equation}
\left|{\cal{E}}_{K}^{(n)}\right|\leq e_{r}^{(2,n)}h^{2},
\end{equation}
where the second-order accuracy is not affected by the dimension.

In order to derive a bound for ${\cal{E}}_{KS}^{(n)}$ let us assume for simplicity
that the crystalline lattice $\Lambda$ is a cube in $n$-dimensions and that
the particles within the cube are unevenly distributed in a low-discrepancy sequence. 
Although in actual SPH applications the computational domains can have a variety of
shapes, the assumption of a cube does not entail a loss of generality. The dual lattice
$\Lambda ^{\star}$ is also an $n$-dimensional cube with finite spectrum
\citep{Meyer16}. As shown in Appendix E, the error bound for the particle
approximation in $n$-dimensions has the form
\begin{equation}
\left|{\cal{E}}_{KS}^{(n)}\right|\leq\left(\frac{2}{\pi}\right)^{n}
\frac{a_{0}B_{n}k^{n}}{n}\sum _{l=0}^{\infty}h^{l}{\tilde e}_{r}^{(l,n)}
\left(\lim _{{\cal{N}}\to\infty}
\frac{1}{\cal{N}}\prod _{s=1}^{n}\sum _{j_{s}=1}^{{\cal{N}}_{s}}
\frac{1}{j_{s}}\right),
\end{equation}
where ${{\cal{N}}_{s}}=[2kh_{s}/\Delta _{s}]$, $h_{s}$ and $\Delta _{s}$ are the
projections of $h$ and $\Delta _{\rm m}$ on the $s$th-axis of an $n$-dimensional
Cartesian system, respectively, $j_{s}$ is the $s$th component of the wave vector
${\bf j}$, and the notation ``$[p]$'' means the largest positive integer less or 
equal to $p$. For low-discrepancy sequences of sample points with ${\cal{N}}\gg 1$,
there will always exist an $n$-dimensional Cartesian system over which the projected
mean distances $\Delta _{s}\approx\Delta _{\rm m}$, and so from Eq. (41) it follows that
${\cal{N}}_{s}\approx (2kh_{s}/{\cal{V}}_{n}^{1/n}){\cal{N}}^{1/n}$. For $N\to\infty$,
$h_{s}\to h\sim {\cal V}_{n}^{1/n}$ and therefore
${{\cal{N}}_{s}}\to\infty$ as ${\cal{N}}\to\infty$. From inequality (28), we find
that
\begin{equation}
\sum _{j_{s}=1}^{{\cal{N}}_{s}}\frac{1}{j_{s}}<\gamma +\ln{{\cal{N}}_{s}}+
\frac{1}{2{\cal{N}}_{s}}.
\end{equation}
Using this into Eq. (43) yields
\begin{eqnarray}
\left|{\cal{E}}_{KS}^{(n)}\right|&\leq&\left(\frac{2}{\pi}\right)^{n}
\frac{a_{0}B_{n}k^{n}}{n}\sum _{l=0}^{\infty}h^{l}{\tilde e}_{r}^{(l,n)}
\left[\lim _{{\cal{N}},{\cal{N}}_{s}\to\infty}
\frac{1}{{\cal{N}}}\prod _{s=1}^{n}\left(\gamma +\ln{{\cal{N}}_{s}}+
\frac{1}{2{\cal{N}}_{s}}\right)\right]\nonumber\\
\nonumber\\
&=&\left(\frac{2}{\pi}\right)^{n}\frac{a_{0}B_{n}k^{n}}{n}\sum _{l=0}^{\infty}
h^{l}{\tilde e}_{r}^{(l,n)}\left[\lim _{{\cal{N}},{\cal{N}}_{s}\to\infty}
\frac{1}{{\cal{N}}}\left(\gamma +\ln{{\cal{N}}_{s}}+\frac{1}{2{\cal{N}}_{s}}\right)^{n}
\right],\nonumber\\
\end{eqnarray}
where
\begin{eqnarray}
\frac{1}{{\cal{N}}}\left(\gamma +\ln{{\cal{N}}_{s}}+\frac{1}{2{\cal{N}}_{s}}\right)^{n}&=&
\frac{{\cal{N}}_{s}^{n}}{{\cal N}}\left(\frac{\gamma}{{\cal{N}}_{s}}+
\frac{\ln{{\cal{N}}_{s}}}{{\cal{N}}_{s}}+\frac{1}{2{\cal{N}}_{s}^{2}}\right)^{n}\nonumber\\
&=&\frac{{\cal{N}}_{s}^{n}}{{\cal N}}\sum _{k=0}^{n}\frac{n!}{2^{k}k!(n-k)!}
\left(\frac{\gamma}{{\cal{N}}_{s}}+
\frac{\ln{{\cal{N}}_{s}}}{{\cal{N}}_{s}}\right)^{n-k}\frac{1}{{\cal{N}}_{s}^{2k}}.\nonumber\\
\end{eqnarray}
Noting that in the limit when ${\cal{N}}_{s}\to\infty$,
the relation $\gamma /{\cal{N}}_{s}+\ln{{\cal{N}}_{s}}/{\cal{N}}_{s}=(\gamma +1)/{\cal{N}}_{s}$
holds, we have that 
\begin{eqnarray}
\lim _{{\cal{N}},{\cal{N}}_{s}\to\infty}\frac{1}{\cal{N}}
\left(\gamma +\ln{{\cal{N}}_{s}}+\frac{1}{2{\cal{N}}_{s}}\right)^{n}&=&
\lim _{{\cal{N}},{\cal{N}}_{s}\to\infty}\frac{{\cal{N}}_{s}^{n}}{\cal{N}}
\left(\frac{\gamma}{{\cal{N}}_{s}}+\frac{\ln{{\cal{N}}_{s}}}{{\cal{N}}_{s}}
+\frac{1}{2{\cal{N}}_{s}^{2}}\right)^{n}\nonumber\\
&=&\lim _{{\cal{N}},{\cal{N}}_{s}\to\infty}\frac{1}{\cal{N}}\sum _{k=0}^{n}
\frac{n!}{k!(n-k)!}\frac{(\gamma +1)^{n-k}}{2^{k}{\cal{N}}_{s}^{k}}.
\nonumber\\
\end{eqnarray}
For low-discrepancy sequences of particles ${\cal{N}}_{s}\propto {\cal{N}}^{1/n}$ and
therefore
\begin{equation}
\lim _{{\cal{N}},{\cal{N}}_{s}\to\infty}\frac{1}{\cal{N}}\sum _{k=0}^{n}
\frac{n!}{k!(n-k)!}\frac{(\gamma +1)^{n-k}}{2^{k}{\cal{N}}_{s}^{k}}\sim
\lim _{{\cal{N}}\to\infty}\left[\frac{(\gamma +1)^{n}}{\cal{N}}+
O\left(\frac{1}{{\cal{N}}^{1+1/n}}\right)\right].
\end{equation}
From the above relations it follows that 
$|{\cal{E}}_{KS}^{(n)}|\to 0$ when ${\cal{N}}\to\infty$. Hence, the asymptotic 
expansion
\begin{equation}
\lim _{{\cal{N}}\to\infty}\frac{1}{\cal{N}}\prod _{s=1}^{n}
\sum _{j_{s}=1}^{{\cal{N}}_{s}}\frac{1}{j_{s}}\leq
\frac{(\gamma +1)^{n}}{\cal{N}}+O\left(\frac{1}{{\cal{N}}^{1+1/n}}\right)
\end{equation}
holds for ${\cal{N}}\to\infty$. For $n=1$, Eq. (49) reduces to Eq. (30) with
the asymptotic bound $O(1/{\cal{N}}^{2})$, while for $n=2$ the asymptotic bound goes as
$O(1/{\cal{N}}^{3/2})$ and for $n=3$ as $O(1/{\cal{N}}^{4/3})$, implying that when
${\cal{N}}\to\infty$ the error $|{\cal{E}}_{KS}^{(n)}|\to 0$ more slowly as the dimension
is increased.
Therefore, the upper bound on the 
particle approximation error takes the form
\begin{equation}
\left|{\cal{E}}_{KS}^{(n)}\right|\leq\left(\frac{2}{\pi}\right)^{n}
\frac{(1+\gamma)^{n}a_{0}B_{n}k^{n}}{n{\cal{N}}}\sum _{l=0}^{\infty}
h^{l}{\tilde e}_{r}^{(l,n)}.
\end{equation}
Retaining only the first two terms in the above summation and adding the bound on 
${{\cal{E}}_{K}^{(n)}}$, as given by inequality (42), the error bound for the full 
SPH estimate of a function is
\begin{eqnarray}
\parallel Sf-If\parallel _{\infty}&\leq&|{{\cal{E}}_{K}^{(n)}}|+|{{\cal{E}}_{KS}^{(n)}}|
\nonumber\\
&\leq&\left(\frac{2}{\pi}\right)^{n}\frac{(1+\gamma)^{n}a_{0}B_{n}k^{n}}
{{n{\cal{N}}}}\left({\tilde e}_{r}^{(0,n)}+h{\tilde e}_{r}^{(1,n)}
+h^{2}{\tilde e}_{r}^{(2,n)}\right)\nonumber\\
&+&h^{2}e_{r}^{(2,n)},
\end{eqnarray}
which is the $n$-dimensional counterpart of the one-dimensional error bound (32). According
to (51), the leading error for the particle approximation in multidimensions goes also as
$1/{\cal{N}}$.

Following steps similar to those described here for the function estimate it can be
demonstrated that the error bounds for the SPH estimate of the gradient obeys a
dependence on the SPH parameters similar to that given by inequality (51).

\section{Discussion}

Using the definition ${{\cal{N}}_{s}}=[2kh_{s}/\Delta _{s}]$ together with Eq. (41), the
error bound (45) can be written in the alternative form
\begin{equation}
\left|{\cal{E}}_{KS}^{(n)}\right|\leq\left(\frac{2}{\pi}\right)^{n}a_{0}
\sum _{l=0}^{\infty}h^{l-n}{\tilde e}_{r}^{(l,n)}\lim _{\Delta _{\rm m},\Delta _{s}\to 0}
\left[\Delta _{\rm m}\left(\gamma -\ln{\Delta _{s}}+\ln (2kh_{s})+\frac{\Delta _{s}}{4kh_{s}}
\right)\right]^{n}.
\end{equation}
For $\Delta _{\rm m}\approx\Delta _{s}$, the term 
$q=-\Delta _{\rm m}\ln\Delta _{s}\approx -\Delta _{\rm m}\ln\Delta _{\rm m}$
provides a measure of the loss of the continuous field information due to the
SPH discretization. As long as
$\Delta _{m}\to 0$, $q\to 0$ and the continuous information is recovered.
This is the essence of the theorem of SPH convergence. On the other hand,
the term $q_{R}=-\Delta _{\rm m}\ln\Delta _{s}$ is a relative measure of the loss of
information since it involves the projections of the mean distance on straight lines.
Since the process of projection works on the way of reducing the information, it is
more convenient to use the SPH interpolation formula on either equidistant or 
low-discrepancy (i.e., quasi-random) sequences of sample points
for which $\Delta _{s}\approx\Delta _{\rm m}$ rather than on randomly disordered
sequences where $\Delta _{s}<\Delta _{\rm m}$.
This point is connected to the average case complexity of multivariate integration
\citep{Wozniakowski91}, where to derive the average case complexity an optimal choice
of the sample points is needed in the computation of multivariate integrals.

\subsection{Approximate partition of unity}

The meaning of partition of unity in SPH is sometimes misunderstood in the literature.
This concept is tightly related to relation (41) and therefore to the volume of the
kernel support. For instance, noting that this relation can be written as
$m_{a}/\rho _{a}=B_{n}k^{n}h^{n}/(n\mathcal{N})+O(h^{n+2})$ and that the volume of the
kernel support is $\mathcal{V}_{n}=B_{n}k^{n}h^{n}/n$, we find that
$m_{a}/\rho _{a}=\mathcal{V}_{n}/\mathcal{N}+O(h^{n+2})$. If according to Lemma 1, we
associate the ratio $m_{a}/\rho _{a}$ to the volume $\Delta V_{a}$ of particle $a$
and sum over all particle within the kernel support, we then recover the relation
$(\sum _{b=1}^{\mathcal{N}}\Delta V_{b})/{\mathcal{V}}_{n}=1+O(h^{n+2})$,
which implies that for finite sizes of the kernel support independently of whether $h$
is fixed or variable, partition of unity can be achieved only approximately to order 
$O(h^{n+2})$. Exact partition of unity can be achieved only when $h\to 0$, which in turn 
will demand that $N\to\infty$, $m\to 0$, and $\mathcal{N}\to\infty$, with 
$\mathcal{N}/N\to 0$, for full consistency. If this requirement is fulfilled, the 
conditions $M_{0,a}=1$ and ${\bf M}_{0,a}^{\prime}={\bf 0}$ will be exactly satisfied.
Moreover, Zhu et al. \cite{Zhu15} quantified numerically the deviation from an exact
partition of unity by evaluating the standard deviation measured in the distributions
of $M_{0,a}$ as a function of ${\mathcal{N}}$ for a low-discrepancy set of particles,
finding that $\sigma (M_{0})\propto {\mathcal{N}}^{-1}$ as is indeed predicted by the
error bound (50). Therefore, the inconsistency in the volume estimate, as measured by
(11), declines as ${\mathcal{N}}$ is increased. As this limit is observed, it was
shown numerically that the SPH estimates of the derivatives converge essentially
at the same rate as the estimate of the function and the approximations become insensitive
to particle disorder \cite{Sigalotti16}. Moreover, as $h\to 0$ the inconsistency
implied by the kernel truncation near a boundary is strongly mitigated and so no special
boundary treatment of the interpolation scheme is needed except for the conditions
demanded by the physics.

\subsection{Numerical experiments}

As a numerical experiment, we consider the SPH reproducibility of the test functions
\begin{equation}
	        f_{\rm 2D}(x,y)=\sin\pi x\sin\pi y,
\end{equation}
defined over the intervals $x\in [0,1]$ and $y\in [0,1]$, and
\begin{equation}
	        f_{\rm 3D}(x,y,z)=\sin\pi x\sin\pi y\sin\pi z,
\end{equation}
defined over the intervals $x\in [0,1]$, $y\in [0,1]$, and $z\in [0,1]$, using the standard
interpolation formula (9) and the scaling relations $\mathcal{N}\propto N^{1/2}$
and $h\propto N^{-1/6}$ \cite{Zhu15}. Out of the family of possible scalings describing
the dependence of $\mathcal{N}$ and $h$ on $N$, we allow $h$ to vary with $N$ as
$h=N^{-1/6}$ to obtain the scaling relations $\mathcal{N}\approx 2.81N^{0.675}$ and
$h\approx 1.29\mathcal{N}^{-0.247}$. The exponents of these scalings are slightly
larger than the reference values of 0.5 and $-1/3$ and were chosen to accommodate
larger numbers of neighbors within the kernel support at the price of having
correspondingly larger values of $h$. A Wendland C$^{4}$ function \cite{Dehnen12}
\begin{equation}
	W(q,h)=\frac{a_{0}}{h^{n}}
        \left\{
	       \begin{array}{cll}
	       (1-q)^{6}\left(1+6q+\frac{35}{3}q^{2}\right) & \mbox{if} & q\leq 1,\\
	       0 & \mbox{if} & q>1,
	       \end{array}
	       \right.
\end{equation}
where $a_{0}=9/\pi$ in 2D ($n=2$), $a_{0}=495/(32\pi)$ in 3D ($n=3$), and
$q=|{\bf x}-{\bf x}^{\prime}|/h$, is employed as the interpolation kernel. In the 2D
case, the test function (53) is represented by filling the unit square plane with
quasi-randomly distributed particles and varying the spatial resolution from $N=25^{2}$
to $N=1000^{2}$. The same irregular pattern is maintained for all resolutions. In the
3D case, the test function (54) is estimated by filling up the unit cube with
particles placed in a similar quasi-random distribution and $N$ varying from
$25^{3}$ to $200^{3}$ particles. The details of the resolution parameters employed in
both sets of calculations are listed in Table 1. In these analyses no boundary
treatments are implemented at the borders of the computational domains.

\begin{table}[htpb]
	\centering
	\caption{Spatial resolution parameters}
	\begin{tabular}{cccccc}
		\hline
		2D & & & 3D & & \\
		\hline
		$N$ & $\mathcal{N}$ & $h$ & $N$ & $\mathcal{N}$ & $h$\\
		\hline
		~~$25^{2}$ & ~~216 & 0.342 & ~$25^{3}$ & ~~1903 & 0.200\\
		~~$50^{2}$ & ~~552 & 0.271 & ~$50^{3}$ & ~~7746 & 0.141\\
		~~$75^{2}$ & ~~955 & 0.237 & ~$75^{3}$ & ~17607 & 0.115\\
		~$100^{2}$ & ~1408 & 0.215 & $100^{3}$ & ~31528 & 0.100\\
		~$125^{2}$ & ~1903 & 0.200 & $125^{3}$ & ~49539 & 0.089\\
		~$150^{2}$ & ~2434 & 0.188 & $150^{3}$ & ~71662 & 0.082\\
		~$175^{2}$ & ~2998 & 0.179 & $175^{3}$ & ~97917 & 0.075\\
		~$200^{2}$ & ~3590 & 0.171 & $200^{3}$ & 128319 & 0.071\\
		~$250^{2}$ & ~4852 & 0.159 & & & \\
		~$300^{2}$ & ~6206 & 0.149 & & & \\
		~$400^{2}$ & ~9151 & 0.136 & & & \\
		~$500^{2}$ & 12368 & 0.126 & & & \\
		~$750^{2}$ & 21381 & 0.110 & & & \\
		$1000^{2}$ & 31528 & 0.100 & & & \\
		\hline
	\end{tabular}
\end{table}

For small values of $\mathcal{N}$ the smoothing length decreases rapidly as $\mathcal{N}$
increases and then more slowly at larger values of $\mathcal{N}$, asymptotically
approaching zero as $\mathcal{N}\to\infty$. The error bound (51) for the test functions 
(53) and (54) reduces to
\begin{equation}
\parallel Sf-If\parallel _{\infty}\leq\frac{36(1+\gamma)^{2}}
{\pi ^{2}\mathcal{N}}\left(1+2\pi h+2\pi ^{2}h^{2}\right)+2\pi ^{2}h^{2},
\end{equation}
and
\begin{equation}
\parallel Sf-If\parallel _{\infty}\leq\frac{165(1+\gamma)^{3}}
{\pi ^{3}\mathcal{N}}\left(1+3\pi h+\frac{9}{2}\pi ^{2}h^{2}\right)+
\frac{9}{2}\pi ^{2}h^{2},
\end{equation}
respectively. Figure 1 shows the error bounds (solid circles and squares) along
with the mean absolute errors (MAEs) between the SPH estimates of the
test functions and their exact values (open circles and squares) calculated
according to
\begin{equation}
\mathcal{E}_{\rm MAE}=\frac{1}{N}\sum _{a=1}^{N}|Sf_{a}-If_{a}|.
\end{equation}
The MAEs and the error bounds (56) and (57) are plotted as a function of $N^{1/2}$
for the 2D runs (circles) and of $N^{1/3}$ for the 3D case (squares). This provides
a better comparison of the errors with resolution in both dimensions. At comparable
values of $N^{1/2}$ and $N^{1/3}$, the 3D particle distributions results in larger
values of $\mathcal{N}$ and smaller smoothing lengths than the 2D case, and
therefore the 3D error bounds are smaller than the 2D counterparts at all resolutions.
As expected, the error bounds tend asymptotically to the MAEs as $N\to\infty$.

As it is well known, the volume estimate of a constant scalar field is not exactly
represented by the standard SPH method because $M_{0}\neq 1$ [see Eq. (11)]. Note
that the discrete form (11) represents the SPH estimate of a constant scalar
field $f=1$. In order to measure the magnitude of this inconsistency and its
dependence on $\mathcal{N}$, we calculate the distributions of $M_{0}$ using
the parameters listed in Table 1 for our 2D and 3D irregularly distributed particles.
The moment $M_{0}$ is expected to follow a peaked distribution around 1 with some
errors for the discrete form (11) to approach the continuous normalization condition
(2). Figures 2 and 3 display the distributions of $M_{0}$ for the 2D and 3D cases,
respectively. For small values of $\mathcal{N}$ the distributions are broad and
their maxima peak at values larger than 1. This means that most particles fall
within a narrow interval whose center deviates from unity and indicates some bias
toward a density overestimate. However, the spread of the distributions is greatly
reduced as $\mathcal{N}$ is increased. As this occurs, the inconsistency of the
density estimates is reduced and the accuracy of the volume estimates increases
in accordance with the $1/\mathcal{N}$-dependence of the error bounds (56) and
(57). As the experiment is repeated with even more neighbors, the distribution of
$M_{0}$ approaches a Dirac-$\delta$ distribution with the error of the volume
estimate decreasing to zero. When this limit is achieved, $C^{0}$-particle consistency
of the SPH method is fully restored. We note that due to the symmetry of the
kernel functions, $C^{1}$-particle consistency is automatically satisfied once
$C^{0}$-consistency is achieved.

As a final remark, we may see that in most practical SPH applications the
distribution of particles is unknown and therefore the overall quality of the density
estimate is difficult to quantify. As was suggested by Zhu et al. \cite{Zhu15}, a
simple procedure to measure the quality of the density estimate is just to
calculate the distribution of $M_{0}$, and possibly all other higher moments up
to the order of accuracy of the kernel function. A recent analysis of several SPH 
schemes has shown
that in the limit of large $\mathcal{N}$ the estimates of a function and its first
derivative converge to the same order independently of the degree of particle
disorder \cite{Sigalotti16}. Therefore, increasing $\mathcal{N}$ regulates the
error observed in standard SPH when passing from a relaxed (regular or quasi-regular)
distribution to an irregular (quasi-random) distribution. This has implications on
relation (39) where exact partition of unity can be achieved only when $h\to 0$,
which in turn demands that $N\to\infty$, $m\to 0$, and $\mathcal{N}\to\infty$. This
is also related to the error carried by the approximation
$\Delta V_{b}\approx m_{b}/\rho _{b}$, which is sensitive to the quality of the
particle distribution. This error is also regulated by increasing $\mathcal{N}$
because as partition of unity is achieved the error of the approximation
$\Delta V_{b}\approx m_{b}/\rho _{b}$ is correspondingly reduced as the volume of
the kernel support $\mathcal{V}_{n}\to\sum _{b=1}^{\mathcal{N}}\Delta V_{b}$.

\subsection{Dependence of the error bounds on the kernel function}

In order to see how the type of the kernel function can influence the convergence
rate, we repeat the experiments of previous section for four different kernel
functions, namely the Lucy's quartic kernel \cite{Lucy77}
\begin{equation}
        W(q,h)=\frac{a_{0}}{h^{n}}
        \left\{
               \begin{array}{cll}
               (1-q)^{3}\left(1+3q\right) & \mbox{if} & q\leq 1,\\
               0 & \mbox{if} & q>1,
               \end{array}
               \right.
\end{equation}
where $a_{0}=5/\pi$ for $n=2$ and $a_{0}=105/16\pi$ for $n=3$, the Monaghan's super
Gaussian kernel \cite{Monaghan92}
\begin{equation}
        W(q,h)=\frac{a_{0}}{h^{n}}
        \left\{
               \begin{array}{cll}
	       \exp(-9q^{2})\left(1+\frac{d}{2}-9q^{2}\right) & \mbox{if} & q\leq 1,\\
               0 & \mbox{if} & q>1,
               \end{array}
               \right.
\end{equation}
where $a_{0}=3^{n}/\pi ^{n/2}$, and the Wendland C$^{2}$ and C$^{6}$ functions given
by
\begin{equation}
        W(q,h)=\frac{a_{0}}{h^{n}}
        \left\{
               \begin{array}{cll}
               (1-q)^{4}\left(1+4q\right) & \mbox{if} & q\leq 1,\\
               0 & \mbox{if} & q>1,
               \end{array}
               \right.
\end{equation}
where $a_{0}=7/\pi$ for $n=2$ and $a_{0}=21/(2\pi)$ for $n=3$, and
\begin{equation}
        W(q,h)=\frac{a_{0}}{h^{n}}
        \left\{
               \begin{array}{cll}
		       (1-q)^{8}\left(1+8q+25q^{2}+32q^{3}\right) & \mbox{if} & q\leq 1,\\
               0 & \mbox{if} & q>1,
               \end{array}
               \right.
\end{equation}
where $a_{0}=78/(7\pi)$ for $n=2$ and $a_{0}=1365/(64\pi)$ for $n=3$, respectively.

Using Eq. (10) the error bounds (56) and (57) for the test functions (53) and (54) can be 
written as 
\begin{equation}
\parallel Sf-If\parallel _{\infty}-\parallel Kf-If\parallel _{\infty}\leq
\parallel Sf-Kf\parallel _{\infty},
\end{equation}
where 
\begin{equation}
\parallel Sf-Kf\parallel _{\infty}\leq\frac{A_{W}^{(2)}(1+\gamma)^{2}}{\pi ^{2}\mathcal{N}}
\left(1+2\pi h+2\pi ^{2}h^{2}\right),
\end{equation}
and
\begin{equation}
\parallel Sf-Kf\parallel _{\infty}\leq\frac{A_{W}^{(3)}(1+\gamma)^{3}}{\pi ^{3}\mathcal{N}}
\left(1+3\pi h+\frac{9}{2}\pi ^{2}h^{2}\right),
\end{equation}
are bounds for the difference between the particle and kernel approximations in 2D
and 3D, respectively. The parameters $A_{W}^{(n)}$ are numerical factors that depend on
the kernel function. Thus, changing the kernel function will affect the convergence rate
of the particle approximation only by a numerical factor. Table 2 lists the values of
$A_{W}^{(n)}$ for the different kernel functions considered. According to these values,
a better choice in 2D would be to use the Lucy's kernel, while in 3D the better choice
would be to use the Wendland C$^{2}$ function. However, as long as $\mathcal{N}$ 
increases, the differences between the various kernels become irrelevant since 
$\parallel Sf-Kf\parallel _{\infty}\to 0$ in the limit $\mathcal{N}\to\infty$
independently of the value of $h$. The mean absolute errors (58) obtained for the SPH 
estimates of the test functions (53) and (54) with the kernels (59)-(62) and the same 
spatial resolution parameters of Table 1 follow the same trends of those plotted in Fig. 1
for the Wendland C$^{4}$ kernel, with differences being less than about 2\% at the lowest 
resolutions and less than 1\% at the highest resolutions. Therefore, as the number of
neighbors increases, the results become independent of the kernel function. However,
when working with large numbers of neighbors we must take care of the fact that most
conventional kernels suffer from a pairing instability, where particles come into close
pairs and become less sensitive to small perturbations within the kernel support
\cite{Dehnen12}. To overcome this difficulty, Wendland-type functions \cite{Wendland95}
are adopted, which have positive Fourier transforms and can support large numbers of
neighbors without inducing a close pairing of particles \cite{Dehnen12}.

\begin{table}[htpb]
        \centering
	\caption{Numerical values of $A_{W}^{(n)}$}
        \begin{tabular}{ccc}
                \hline
                Kernel & $A_{2}^{(n)}$ & $A_{2}^{(n)}$\\
                \hline
		Lucy & ~20 & 210\\
		Super Gaussian & ~72 & 720/$\sqrt{\pi}$\\
		Wendland C$^{2}$ & ~28 & 112\\
		Wendland C$^{4}$ & ~36 & 165\\
		Wendland C$^{6}$ & 156 & 455/2\\
                \hline
        \end{tabular}
\end{table}

According to the error bound (51), the accuracy of the particle approximation approaches
that of the kernel approximation only when $\mathcal{N}\to\infty$ regardless of the value
of $h$, while full convergence to the exact solution can be obtained only when also 
$h\to 0$. On the other hand, very small smoothing lengths with a small number of neighbors
($\mathcal{N}<100$), as employed in most conventional SPH calculations, is not enough
to guarantee convergence since in the limit $h\to 0$, an irreducible zeroth-order error
term proportional to $1/\mathcal{N}$ will still be present.

\subsection{SPH approximation of the Dirac-$\delta$ distribution}

There is a point which has not been addressed directly in the SPH literature. This point
is concerned with the SPH estimate of the Dirac-$\delta$ distribution. In principle, we
may use the Poisson summation given by Eq. (15) to evaluate the convergence of the SPH 
estimate of the distribution $\delta (x_{a}-x_{0})$, namely
\begin{eqnarray}
\delta (x_{a}-x_{0})&=&\int _{\Omega _{1}}\delta (x_{b}-x_{0})W(|x_{a}-x_{b}|,h)dx_{b}
\nonumber\\
&+&2\sum _{j=1}^{\infty}\int _{\Omega _{1}}\delta (x_{b}-x_{0})W(|x_{a}-x_{b}|,h)
\cos (2\pi jb)dx_{b},
\end{eqnarray}
where $x_{0}\in{\mathbb{R}}^{+}$. After using the sampling property of the Dirac-$\delta$
function, Eq. (66) reduces to
\begin{equation}
        \delta (x_{a}-x_{0})=
        \left\{
               \begin{array}{cll}
               W(|x_{a}-x_{0}|,h)\left[1+2\sum _{j=1}^{\infty}\cos (2\pi jb_{0})\right] & 
               \mbox{if} & x_{0}\in\Omega _{1},\\
               0 & \mbox{if} & x_{0}\notin\Omega _{1},
               \end{array}
               \right.
\end{equation}
where $b_{0}\in\mathbb{N}$ is a label associated to position $x_{0}$. Regularization
of the summation in Eq. (67) in the sense of the criterion (B.11) in Appendix B yields
\begin{equation}
\sum _{j=1}^{\infty}\cos (2\pi jb_{0})=\lim _{\mathcal{N}\to\infty}
\sum _{j=1}^{\mathcal{N}}\cos (2\pi jb_{0})=\lim _{\mathcal{N}\to\infty}
{\rm Re}\left(\sum _{j=1}^{\mathcal{N}}\exp(i2\pi jb_{0})\right),
\end{equation}
where $i=\sqrt{-1}$. If $b_{0}\in\mathbb{N}$, we have that
\begin{equation}
\lim _{\mathcal{N}\to\infty}{\rm Re}\left(\sum _{j=1}^{\mathcal{N}}
\exp(i2\pi jb_{0})\right)=\infty,
\end{equation}
while, if $b_{0}\in\mathbb{R}\setminus\mathbb{N}$, then any machine representation of
$b_{0}$ is of the form $s+r/q$, with $s,r,q\in\mathbb{N}$ and $r<q$. Then,
\begin{equation}
{\rm Re}\left(\sum _{j=1}^{\mathcal{N}}\exp(i2\pi jb_{0})\right)=
{\rm Re}\left\{\frac{[\exp(i2\pi\mathcal{N}r/q)-1]}{[\exp(i2\pi r/q)-1]}
\exp(i2\pi r/q)\right\}=0,
\end{equation}
if $\mathcal{N}\equiv 0$ ${\rm mod}q$. When $\mathcal{N}\not\equiv 0$ ${\rm mod}q$,
the series is different from zero and has a finite number of terms. This suggests
that the SPH estimate of the Dirac-$\delta$ function has an oscillatory behavior as
a function of $\mathcal{N}$. On the other hand, if the source pointed function is
localized on a particle, the interpolation will diverge linearly from the exact
solution even for $\mathcal{N}\gg 1$. To overcome these difficulties, the
Dirac-$\delta$ distribution is replaced by a regularized smooth function
\begin{equation}
\delta (x-x_{0})=\lim _{\sigma\to 0^{+}}\eta _{\sigma}(x-x_{0}),
\end{equation}
where $\eta _{\sigma}(x)$ is sometimes called a nascent $\delta$ function, which has
the following scaling properties
\begin{equation}
\eta _{\sigma}({\bf x})=\frac{1}{\sigma ^{n}}\eta\left(\frac{{\bf x}}{\sigma}\right),
\end{equation}
where $\sigma$ is the bandwidth of $\eta _{\sigma}(x)$ and, as before, $n$ denotes the 
spatial dimension. For simplicity, let us consider
a Gaussian distribution with standard deviation $\sigma$ so that
\begin{equation}
\delta (x-x_{0})=\lim _{\sigma\to 0^{+}}\frac{1}{\sqrt{\pi}\sigma}
\exp\left[-\frac{(x-x_{0})^{2}}{\sigma ^{2}}\right],
\end{equation}
and perform the analysis in one-space dimension ($n=1$). A similar procedure follows
in two- and three-dimensions. We start by expanding an arbitrary function $f(x)$ in
Taylor series about $x=x_{0}$ such that
\begin{equation}
f(x)=f(x_{0})+\sum _{k=1}^{\infty}\frac{1}{k!}f^{(k)}(x_{0})(x-x_{0})^{k}.
\end{equation}
According to the sampling property of the Dirac-$\delta$ distribution, we have that
\begin{equation}
f(x_{0})=\lim _{\sigma\to 0^{+}}\int _{-\infty}^{\infty}f(x)\frac{1}{\sqrt{\pi}\sigma}
\exp\left[-\frac{(x-x_{0})^{2}}{\sigma ^{2}}\right]dx.
\end{equation}
Substitution of the expansion (74) into Eq. (75) gives for $f(x_{0})$ the expression
\begin{eqnarray}
f(x_{0})&=&\lim _{\sigma\to 0^{+}}\left\{f(x_{0})\frac{1}{\sqrt{\pi}\sigma}
\int _{-\infty}^{\infty}\exp\left[-\frac{(x-x_{0})^{2}}{\sigma ^{2}}\right]dx\right.
\nonumber\\
&+&\left.\frac{1}{\sqrt{\pi}\sigma}\sum _{k=0}^{\infty}\frac{1}{(2k+1)!}f^{(2k+1)}(x_{0})
\int _{-\infty}^{\infty}(x-x_{0})^{2k+1}
\exp\left[-\frac{(x-x_{0})^{2}}{\sigma ^{2}}\right]dx\right.\nonumber\\
&+&\left.\frac{1}{\sqrt{\pi}\sigma}\sum _{k=1}^{\infty}\frac{1}{(2k)!}f^{(2k)}(x_{0})
\int _{-\infty}^{\infty}(x-x_{0})^{2k}
\exp\left[-\frac{(x-x_{0})^{2}}{\sigma ^{2}}\right]dx\right\},
\end{eqnarray}
where the first integral is equal to one by the normalization condition of the Gaussian
distribution for $\forall\sigma\in\mathbb{R}^{+}$, the second term vanishes because the
odd moments of the Gaussian distribution are exactly zero by symmetry, and the last term 
survives because the even moments obey the relation
\begin{equation}
\int _{-\infty}^{\infty}(x-x_{0})^{2k}
\exp\left[-\frac{(x-x_{0})^{2}}{\sigma ^{2}}\right]dx=\sigma ^{2k+1}\Gamma
\left(k+\frac{1}{2}\right),
\end{equation}
where $\Gamma (z)$ is the Gamma function with $z>0$. Therefore,
\begin{equation}
f(x_{0})=f(x_{0})+\lim _{\sigma\to 0^{+}}\frac{\sigma ^{2k}}{\sqrt{\pi}}
\sum _{k=1}^{\infty}\frac{1}{(2k)!}f^{(2k)}(x_{0})\Gamma\left(k+\frac{1}{2}\right),
\end{equation}
which demonstrates the validity of Eq. (71). Retaining only the first term in the summation, 
it is easy to see that in the limiting process to zero, the nascent Dirac-$\delta$ function is 
a second-order approximation to the Dirac-$\delta$ distribution since
\begin{equation}
f(x_{0})=f(x_{0})+\lim _{\sigma\to 0^{+}}\left[\frac{1}{4}f^{(2)}(x_{0})\sigma ^{2}
+O(\sigma ^{4})\right].
\end{equation}
The same result is also obtained in $n$-dimensions where
\begin{equation}
f({\bf x}_{0})=f({\bf x}_{0})+\lim _{\sigma\to 0^{+}}\left[\frac{1}{4}
\nabla ^{2}f({\bf x}_{0})\sigma ^{2}+O(\sigma ^{4})\right].
\end{equation}
As in Section 5.2, the error bound (51) for the SPH approximation of the Dirac-$\delta$
distribution in two-dimensions normalized to $1/(\pi\sigma ^{2})$ can be calculated to 
give
\begin{equation}
\parallel Sf-If\parallel _{\infty}\leq\frac{36(1+\gamma)^{2}}{\pi ^{2}\mathcal{N}}
\left(1+\frac{2\sqrt{2}}{\sqrt{e}}\frac{h}{\sigma}+4\frac{h^{2}}{\sigma ^{2}}\right)
+4\frac{h^{2}}{\sigma ^{2}},
\end{equation}
while after normalization by the factor $1/(\pi ^{3/2}\sigma ^{3})$, the error bound in
three-dimensions reads as follows
\begin{equation}
\parallel Sf-If\parallel _{\infty}\leq\frac{165(1+\gamma)^{3}}{\pi ^{3}\mathcal{N}}
\left(1+3\sqrt{2}\frac{h}{\sigma}+9\frac{h^{2}}{\sigma ^{2}}\right)
+9\frac{h^{2}}{\sigma ^{2}}.
\end{equation}
From expressions (81) and (82) we may see immediately that regardless of the value of
$h/\sigma$, the error due to the particle approximation decays as $\mathcal{N}^{-1}$,
meaning that for large $\mathcal{N}$ this error is smaller than the kernel approximation
error given by the last term in the right-hand side of expressions (81) and (82). Therefore, 
convergence to the Dirac-$\delta$ distribution can only be obtained if, in addition, $h\to 0$ 
faster than $\sigma$. Therefore, only in the limit when $\mathcal{N}\to\infty$, $h\to 0$, and
$\sigma\to 0$, with $h/\sigma\to 0$, complete convergence is achieved. Since the number
of particles within an $n$-dimensional sphere of radius $\sigma$ around the maximum of
the Gaussian distribution is
\begin{equation}
{\mathcal{N}}_{\sigma}=\left(\frac{\sigma}{kh}\right)^{n}\mathcal{N},
\end{equation}
we have that for $k=1$
\begin{equation}
\frac{\mathcal{N}}{{\mathcal{N}}_{\sigma}}=\frac{h^{2}}{\sigma ^{2}},
\end{equation}
for $n=2$, and
\begin{equation}
\frac{\mathcal{N}}{{\mathcal{N}}_{\sigma}}=\frac{h^{3}}{\sigma ^{3}},
\end{equation}
for $n=3$. Replacing these expressions in Eqs. (81) and (82), we may see that claiming
that $h/\sigma\ll 1$ for strict convergence is equivalent to the requirement that 
$\mathcal{N}/{{\mathcal{N}}_{\sigma}}\ll 1$, i.e., the number of particles within a
sphere of radius $\sigma$ must always be larger than the number of neighbors within
the support of the interpolating kernel.

\begin{table}[htpb]
        \centering
        \caption{Spatial resolution parameters for the SPH Dirac-$\delta$ estimation}
        \begin{tabular}{ccccc}
                \hline
		   & & $(h/\sigma)$ & & \\                
		\hline
		$N$ & $\mathcal{N}$ & $\sigma =0.1$ & $\sigma =0.01$ & $\sigma =0.001$ \\ 
                \hline
		2D & & & & \\
		\hline
		~~$50^{2}$ & ~~552 & 2.71 & 27.1 & 271 \\
		~$125^{2}$ & ~1903 & 2.00 & 20.0 & 200 \\
		~$200^{2}$ & ~3590 & 1.71 & 17.1 & 171 \\
                ~$400^{2}$ & ~9151 & 1.36 & 13.6 & 136 \\
                $1000^{2}$ & 31528 & 1.00 & 10.0 & 100 \\
		\hline
		3D & & & & \\
		\hline
		~$50^{3}$ & ~~7746 & 1.41 & 14.1 & 141\\
		$125^{3}$ & ~49539 & 0.89 & ~8.9 & ~89\\
		$200^{3}$ & 128319 & 0.71 & ~7.1 & ~71\\
		\hline
        \end{tabular}
\end{table}

Table 3 lists the values of $h/\sigma$ for the 2D and 3D experiments when $\sigma$ takes
values of 0.1, 0.01, and 0.001. For the 2D case, we may see that for all the resolutions
tried in this paper $h/\sigma\geq 1$, implying that at these spatial resolutions 
convergence will never be achieved as was confirmed numerically for the case
$f(x,y)=\delta (x-x_{0})\delta (y-y_{0})$, with $x_{0}=0.5$ and defined over the intervals
$x\in [0,1]$ and $y\in [0,1]$. In the 3D case, for a function
$f(x,y,z)=\delta (x-x_{0})\delta (y-y_{0})\delta(z-z_{0})$, with $x_{0}=0.5$ and defined
over the intervals $x\in [0,1]$, $y\in [0,1]$, and $z\in [0,1]$, only when
$\sigma =0.1$ is the ratio $h/\sigma <1$ for the highest resolutions. However, at these
resolutions as $\sigma\to 0$, the ratio $h/\sigma$ becomes larger and larger, implying
that convergence to the Dirac-$\delta$ distribution would demand increasing both $N$ and 
$\mathcal{N}$ such that in the limiting process $h/\sigma\to 0$. As expected the numerical
simulations show that for the resolutions of Table 3 the MAEs oscillate as predicted by 
Eq. (67). 

\subsection{Rounding error analysis of the SPH interpolation formula}

We now consider the error made when the particle approximation given by Eq. (9) is
computed in floating-point. The summation (9) is an extended sequence of additions
and products, where $f_{b}$, $W_{ab}$, and $\Delta V_{b}$ are standard floating-point
numbers. In this sequence, the products are computed first and then added in the order
in which they are written. The floating-point machine number that corresponds to the
real value of the particle estimate $f_{a}$ is denoted by \cite{Wilkinson94}
\begin{eqnarray}
{\rm fl}(f_{a})&=&{\rm fl}\left(\sum_{b=1}^{\mathcal{N}}f_{b}W_{ab}\Delta V_{b}\right)
\nonumber\\
&=&{\rm fl}\left(f_{1}W_{a1}\Delta V_{1}+f_{2}W_{a2}\Delta V_{2}+\cdots +
f_{\mathcal{N}}W_{a\mathcal{N}}\Delta V_{\mathcal{N}}\right).
\end{eqnarray}
In order to evaluate Eq. (86) we must define the following quantities recursively
\begin{eqnarray}
s_{1}&=&{\rm fl}\left(f_{1}W_{a1}\Delta V_{1}\right)=f_{1}W_{a1}\Delta V_{1}(1+\eta _{1}),
\nonumber\\
s_{2}&=&{\rm fl}\left(s_{1}+{\rm fl}\left(f_{2}W_{a2}\Delta V_{2}\right)\right)=
{\rm fl}\left(s_{1}+f_{2}W_{a2}\Delta V_{2}(1+\eta _{2})\right)\nonumber\\
&=&f_{1}W_{a1}\Delta V_{1}(1+\eta _{1})(1+\delta _{2})+
f_{1}W_{a2}\Delta V_{2}(1+\eta _{2})(1+\delta _{2}),\nonumber\\
s_{3}&=&{\rm fl}\left(s_{2}+{\rm fl}\left(f_{3}W_{a3}\Delta V_{3}\right)\right)=
{\rm fl}\left(s_{2}+f_{3}W_{a3}\Delta V_{3}(1+\eta _{3})\right)\nonumber\\
&=&f_{1}W_{a1}\Delta V_{1}(1+\eta _{1})(1+\delta _{2})(1+\delta _{3})+\nonumber\\
&+&f_{2}W_{a2}\Delta V_{2}(1+\eta _{2})(1+\delta _{2})(1+\delta _{3})+\nonumber\\
&+&f_{3}W_{a3}\Delta V_{3}(1+\eta _{3})(1+\delta _{3}),
\end{eqnarray}
and so on. By induction, we have for $b=\mathcal{N}$ that
\begin{eqnarray}
s_{\mathcal{N}}&=&f_{1}W_{a1}\Delta V_{1}(1+\eta _{1})(1+\delta _{2})(1+\delta _{3})+
\cdots +(1+\delta _{\mathcal{N}})+\nonumber\\
&+&f_{2}W_{a2}\Delta V_{2}(1+\eta _{2})(1+\delta _{2})(1+\delta _{3})+\cdots +
(1+\delta _{\mathcal{N}})+\nonumber\\
&+&\cdots +f_{\mathcal{N}}W_{a\mathcal{N}}\Delta V_{\mathcal{N}}(1+\eta _{\mathcal{N}})
(1+\delta _{\mathcal{N}}),
\end{eqnarray}
where $|\eta _{b}|\leq u$ and $|\delta _{b}|\leq u$, with $\delta _{1}=0$. Here, $u$
is the unit round-off error defined as $u=\epsilon /2$, where $\epsilon =\beta ^{-(t-1)}$
is the so-called machine epsilon. In IEEE standard double precision, $\beta =2$ and $t=53$ 
so that $\epsilon =2^{-52}=2.220446\times 10^{-16}$ and $u=1.110223\times 10^{-16}$.
From Eqs. (87) and (88) it follows that
\begin{equation}
{\rm fl}(f_{a})=\sum _{b=1}^{\mathcal{N}}f_{b}W_{ab}\Delta V_{b}(1+\eta _{b})
\prod _{c=b}^{\mathcal{N}}(1+\delta _{c}).
\end{equation}
For $b=1$, the number of coefficients of $f_{1}W_{a1}\Delta V_{1}$ is equal to $\mathcal{N}$ 
since $\delta _{1}=0$, while for $b>1$, the number of coefficients of $f_{b}W_{ab}\Delta V_{b}$ 
is just $\mathcal{N}-b+2$. Since $|\eta _{b}|\leq u$ and $|\delta _{c}|\leq u$, we may write
that
\begin{equation}
(1+\eta _{b})\prod _{c=b}^{\mathcal{N}}(1+\delta _{c})
\approx (1+\eta _{b})(1+\delta)^{\mathcal{N}-1}\approx (1+\delta)^{\mathcal{N}}.
\end{equation}
Now, expanding in Taylor series $(1+\delta)^{\mathcal{N}}$ about $\delta =0$ and retaining
terms up to the first order, we have that
\begin{equation}
(1+\delta)^{\mathcal{N}}\approx 1+\mathcal{N}\delta,
\end{equation}
so that Eq. (89) can be written in the much simpler form
\begin{equation}
{\rm fl}(f_{a})=(1+\mathcal{N}\delta)\sum _{b=1}^{\mathcal{N}}f_{b}W_{ab}\Delta V_{b}
=(1+\mathcal{N}\delta)f_{a},
\end{equation}
with $|\delta|\leq\epsilon /2$. Therefore, a bound on the forward error involved in the
operation is
\begin{eqnarray}
\left|{\rm fl}\left(\sum_{b=1}^{\mathcal{N}}f_{b}W_{ab}\Delta V_{b}\right)-
\sum_{b=1}^{\mathcal{N}}f_{b}W_{ab}\Delta V_{b}\right|&\leq&
\left|\mathcal{N}\delta\sum _{b=1}^{\mathcal{N}}f_{b}W_{ab}\Delta V_{b}\right|
\nonumber\\
&\leq&\mathcal{N}|\delta|\sum _{b=1}^{\mathcal{N}}|f_{b}|W_{ab}\Delta V_{b}\nonumber\\
&\leq&\frac{1}{2}\mathcal{N}\epsilon |f_{a}|.
\end{eqnarray}
From the last inequality in Eq. (93) it follows that the particle estimate $f_{a}$ is
approximated by a floating number ${\rm fl}(f_{a})$ with a relative error no larger
than ${\mathcal{N}\epsilon /2}$. In IEEE standard double precision, this error is
$\leq 1.110223\times 10^{-16}\mathcal{N}$ and becomes larger with larger number of
neighbors.

\section{Conclusions}

The consistency and convergence of the smoothed particle hydrodynamics (SPH) interpolation
formula for the estimate of a function was investigated by analytical means. Because of
the widespread use of SPH in science and engineering, the issue of SPH consistency has
become a very hot and important topic of research. The method employed to derive the
explicit functional dependence of the error bounds on the SPH interpolation parameters
for the particle approximation of a function is based on the Poisson summation formula
for kernels with a locally finite support. The results of the analysis not only clarify
the issue of SPH consistency, but also permit assessing the accuracy of the standard SPH
interpolation formula which has been thought to be a non-trivial problem.

The advantage of using the Poisson summation formula is that it enables the simultaneous
treatment of both the kernel and particle approximation errors from which new
consistency integral relations for the particle estimate follow as the cosine Fourier
transform of the kernel consistency relations.
The functional dependence of the error bounds on the SPH parameters, namely the smoothing
length, $h$, and the number of neighbors, ${\mathcal N}$, within the kernel support is
derived explicitly from which consistency conditions arise. In
particular, as long as ${\mathcal N}\to\infty$, the particle approximation converges to
the kernel approximation independently of $h$ provided that the particle mass scales with
$h$ as $m\propto h^{\beta}$, with $\beta >n$, where $n$ is the spatial dimension. This
implies that as $h\to 0$, the joint limit $m\to 0$, ${\mathcal N}\to\infty$, and
$N\to\infty$ is necessary to restore complete consistency, where $N$ is the total number
of particles. The requirement that $m\to 0$ as $h\to 0$ leads to the scaling
${\mathcal N}\propto h^{n-\beta}$ \cite{Zhu15}. In addition, for finite values of 
${\mathcal N}$ a dominant error of the form $(\ln {\mathcal N})^{n}/{\mathcal N}$ emerges 
from the present analysis, as was first conjectured by Monaghan \cite{Monaghan85} based on 
the similarity between the SPH and the quasi-Monte Carlo estimates. For ${\mathcal{N}}\gg 1$, 
the present analysis predicts that the error of the SPH approximation declines as 
${\mathcal{N}}^{-1}$ independently of the dimension, guaranteeing approximate partition of 
unity of the kernel volume.

In the light of the above results, the Poisson summation formula appears to be a
powerful tool for the error analysis of particle methods involving the evaluation
of quadratures, as is also the case of the element-free Galerkin (EFG and GEFG), the
reproducing kernel particle (RKPM and GRKPM), the moving least squares (MLSM and GMLSM),
the Monte Carlo, and the quasi-Monte Carlo schemes among others.
On the other hand, application of the present method to the analysis of the fluid-dynamics
SPH equations will allow to formally assess the accuracy and convergence of current
SPH simulations by analytical means.

\begin{appendices}
\setcounter{equation}{0}
\renewcommand\theequation{A.\arabic{equation}}
\section{Error bound for the kernel approximation in one dimension}

Using the kernel normalization condition (2) and recalling that the first
moment ($l=1$) of the kernel vanishes identically, the kernel approximation
of $f(x)$ at the position of particle $a$ follows from the first summation
on the right side of Eq. (18) as
\begin{eqnarray}
\langle f(x_{a})\rangle &=&f(x_{a})\nonumber\\
&+&\frac{1}{2}f^{(2)}(x_{a})\int _{\Omega _{1}}(x_{b}-x_{a})^{2}
W(|x_{a}-x_{b}|,h)dx_{b},
\end{eqnarray}
where only terms up to $l=2$ have been retained. Using the Cauchy-Schwarz 
inequality it follows that
\begin{eqnarray}
|{\cal{E}}_{K}|&=&\frac{1}{2}\left|f^{(2)}(x_{a})
\int _{\Omega _{1}}(x_{b}-x_{a})^{2}W(|x_{a}-x_{b}|,h)dx_{b}\right|
\nonumber\\
&\leq&\frac{1}{2}\left|f^{(2)}(x_{a})\right|
\int _{\Omega _{1}}\left|(x_{b}-x_{a})^{2}\right|W(|x_{a}-x_{b}|,h)dx_{b}.
\end{eqnarray}
Noting that $\left|(x_{b}-x_{a})^{2}\right|\leq k^{2}h^{2}$ and defining
\begin{equation}
e_{r}^{(2)}=\frac{k^{2}}{2}\sup _{\xi\in\Omega _{1}}
\left|f^{(2)}(\xi)\right|\geq\frac{k^{2}}{2}\left|f^{(2)}(\xi)\right|,
\end{equation}
the bound is
\begin{equation}
|{\cal{E}}_{K}|\leq e_{r}^{(2)}h^{2}\int _{\Omega _{1}}
W(|x_{a}-x_{b}|,h)dx_{b}=e_{r}^{(2)}h^{2},
\end{equation}
which is second-order in $h$.

\setcounter{equation}{0}
\renewcommand\theequation{B.\arabic{equation}}
\section{Error bound for the particle approximation in one dimension}

The double summation in the second term on the right side of Eq. (18) gives
the error due to the particle discretization and represents the difference
${\cal{E}}_{KS}=f_{a}-\langle f(x_{a})\rangle$ between the particle
approximation and the kernel estimate of $f(x)$ evaluated at the position
of particle $a$. The steps involved in the derivation of inequality (27) are
described here starting from
\begin{equation}
{\cal{E}}_{KS}=2\sum _{j=1}^{\infty}\sum _{l=0}^{\infty}\frac{f^{(l)}(x_{a})}{l!}
\int _{\Omega _{1}}(x_{b}-x_{a})^{l}W(|x_{a}-x_{b}|,h)\cos (2\pi jb)dx_{b}.
\end{equation}
To obtain a bound on this error term first define
\begin{equation}
{\tilde e}_{r}^{(l)}=\frac{k^{l}}{l!}\sup _{\xi\in\Omega _{1}}
\left|f^{(l)}(\xi)\right|\geq\frac{k^{l}}{l!}
\left|f^{(l)}(\xi)\right|,
\end{equation}
and $|(x_{b}-x_{a})^{l}|\leq k^{l}h^{l}$. Moreover, since any suitable kernel
function achieves a maximum value at the position of the observation point,
i.e., $\max\{W|(x_{a}-x_{b}|,h)\}=W(0,h)=a_{0}/h$ when $x_{b}=x_{a}$, it 
follows that $W(|x_{a}-x_{b}|,h)\leq a_{0}/h$ for any $x_{b}\in {\rm supp}(W)$,
where $a_{0}$ is a positive constant. Therefore,
\begin{eqnarray}
\left|{\cal{E}}_{KS}\right|&=&2\left|\sum _{j=1}^{\infty}\sum _{l=0}^{\infty}
\frac{f^{(l)}(x_{a})}{l!}
\int _{\Omega _{1}}(x_{b}-x_{a})^{l}W(|x_{a}-x_{b}|,h)
\cos (2\pi jb)dx_{b}\right|\nonumber\\
&\leq&2\sum _{j=1}^{\infty}\sum _{l=0}^{\infty}\frac{1}{l!}
\left|f^{(l)}(x_{a})\right|
\left|\int _{\Omega _{1}}(x_{b}-x_{a})^{l}W(|x_{a}-x_{b}|,h)
\cos (2\pi jb)dx_{b}\right|.\nonumber\\
\end{eqnarray}
Let $x_{b,\max}$ be a point within $\Omega _{1}$ such that
$(x_{b,\max}-x_{a})^{l}W(|x_{a}-x_{b,\max}|,h)=\max\{(x_{b}-x_{a})^{l}W(|x_{a}-x_{b}|,h)\}$,
$\forall x_{b}\in\Omega _{1}$. Using the Cauchy-Schwarz inequality in (B.3), the error 
bound follows as
\begin{eqnarray}
\left|{\cal{E}}_{KS}\right|&\leq&2\sum _{j=1}^{\infty}\sum _{l=0}^{\infty}\frac{1}{l!}
\left|f^{(l)}(x_{a})\right|\left|(x_{b,\max}-x_{a})^{l}\right|W(|x_{a}-x_{b,\max}|,h)
\left|\int _{\Omega _{1}}\cos (2\pi jb)dx_{b}\right|\nonumber\\
&\leq&2a_{0}\sum _{l=0}^{\infty}h^{l-1}{\tilde e}_{r}^{(l)}\sum _{j=1}^{\infty}
\left|\int _{\Omega _{1}}\cos (2\pi jb)dx_{b}\right|,
\end{eqnarray}
where relation (B.2) and the definitions $|(x_{b,\max}-x_{a})^{l}|\leq k^{l}h^{l}$ and
$W(|x_{a}-x_{b,\max}|,h)\leq a_{0}/h$ have been used. To evaluate the cosine integral
first expand $b=b(x_{b})$ about $x_{a}$ to produce the linear mapping
\begin{equation}
b(x_{b})=b(x_{a})+\left(\frac{db}{dx_{b}}\right)_{x_{b}=x_{a}}(x_{b}-x_{a})+
O[(x_{b}-x_{a})^{2}],
\end{equation}
where according to relations (16) and (21) 
\begin{equation}
\left(\frac{db}{dx_{b}}\right)_{x_{b}=x_{a}}=\frac{\rho (x_{a})}{m(x_{a},h)}=
\frac{{\cal{N}}(x_{a},h)}{2kh}.
\end{equation}
Note that if $\cos (2\pi jb)$ varies rapidly within the domain $\Omega _{1}$ the
above integral vanishes, which is a necessary requirement to test its convergence
to zero when ${\cal{N}}\gg 1$. Replacing relation (B.6) into expansion (B.5), the 
cosine integral becomes
\begin{eqnarray}
\int _{\Omega _{1}}\cos (2\pi jb)dx_{b}&=&\cos [2\pi jb(x_{a})]\int _{x_{a}-kh}^{x_{a}+kh}
\cos\left[\frac{2\pi}{P}x_{ba}\right]dx_{b}\nonumber\\
&-&\sin [2\pi jb(x_{a})]\int _{x_{a}-kh}^{x_{a}+kh}
\sin\left[\frac{2\pi}{P}x_{ba}\right]dx_{b},
\end{eqnarray}
where $P=2hk/j{\cal{N}}(\xi ,h)$ and $x_{ba}=x_{b}-x_{a}$. The sine integral vanishes
identically, while the cosine integral has a maximum value equal to $P/\pi$. Therefore,
\begin{equation}
\left|\int _{x_{a}-kh}^{x_{a}+kh}\cos\left[\frac{2\pi}{P}x_{ba}\right]dx_{b}\right|
\leq\frac{P}{\pi},
\end{equation}
from which it follows that
\begin{equation}
\left|\int _{\Omega _{1}}\cos (2\pi jb)dx_{b}\right|\leq
\left|\cos [2\pi jb(x_{a})]\right|\frac{2kh}{\pi j{\cal{N}}(x_{a},h)}
\leq\frac{2kh}{\pi j{\cal{N}}(x_{a},h)}.
\end{equation}
Replacing this bound into (B.4) leads to inequality (27)
\begin{equation}
\left|{\cal{E}_{KS}}\right|\leq\frac{4}{\pi}a_{0}k\sum _{l=0}^{\infty}
h^{l}{\tilde e}_{r}^{(l)}
\left(\lim _{{\cal{N}}(x_{a},h)\to\infty}\frac{1}{{\cal{N}}(x_{a},h)}
\sum _{j=1}^{{\cal{N}}(x_{a},h)}\frac{1}{j}\right).
\end{equation}
In writing inequality (B.10) the following must be noticed. First, returning to Eq. (13)
we recall that the function $\phi (b)$ in the leftmost sum has locally finite support
and therefore the correspondence in Eq. (14) holds, where the integer $b\in [1,{\cal{N}}]$.
That is, only the ${\cal{N}}$ sample points within the support of $\phi (b)$ will
actually contribute to the sum. According to the Nyquist-Shannon sampling theorem a
sufficient condition for a discrete sequence of samples to capture all the information
from the continuous function $\phi (b)$ is that $j_{\rm max}={\cal{N}}$. Thus,
regularization would demand writing
\begin{equation}
\sum _{j=1}^{\infty}\to\lim _{A\to\infty}\sum _{j=1}^{A}\leq
\lim _{{\cal{N}}\to\infty}\sum _{j=1}^{{\cal{N}}}, 
\end{equation}
for $A\in\mathbb{N}$ and ${\cal{N}}\geq A$. For finite values of ${\cal{N}}$, $j$, and
$h$, the last term in inequality (B.9) will always be positive. Since regularization
demands that ${\cal{N}}\geq A$, the inequality in (B.11) holds and so the term between
parentheses in (B.10) will always converge to zero when ${\cal{N}}\to\infty$.

\setcounter{equation}{0}
\renewcommand\theequation{C.\arabic{equation}}
\section{One-dimensional convergence analysis for equidistant particles}

The error analysis developed in Section 3 can be tested for equidistant particles
with spacing $\Delta$ over an infinite line so that $m/\rho =\Delta$. If, in addition,
we consider the interpolation of the linear function $f(x)=\alpha +\beta x$ with
the use of the Gaussian kernel
\begin{equation}
W_{G}(x,h)=\frac{1}{\sqrt{\pi}h}\exp\left(-\frac{x^{2}}{h^{2}}\right),
\end{equation}
we can estimate the error in the SPH summation interpolant by calculating all terms 
in the series expansion on the right side of Eq. (13). For equidistant particles, 
Eq. (17) reduces to $x_{b}=b\Delta$. The SPH interpolation formula (9) gives, at 
$x_{a}=a\Delta$, the following approximation for $f(x)$
\begin{equation}
f_{a}=\Delta\sum _{b=-\infty}^{\infty}(\alpha +\beta b\Delta)W(|a\Delta -b\Delta |,h).
\end{equation}
If we shift the origin to the point $x_{a}=a\Delta$, make the change of variable
$b=q/\Delta$, and use the Poisson summation formula (13), the SPH approximation
of the linear function becomes
\begin{equation}
f_{a}=(\alpha +\beta a\Delta)\left[1+2\sum _{j=1}^{\infty}
\int _{-\infty}^{\infty}\cos\left(2\pi j\frac{q}{\Delta}\right)W_{G}(q,h)dq\right].
\end{equation}
Using the Gaussian kernel (C.1), Eq. (C.3) reduces to 
\begin{equation}
f_{a}=(\alpha +\beta a\Delta)\left[1+2\sum _{j=1}^{\infty}
\exp\left(-\frac{\pi ^{2}j^{2}h^{2}}{\Delta ^{2}}\right)\right].
\end{equation}
It is easy to show that the series expansion on the right side of Eq. (C.4) is
absolutely convergent for all values of $h$ and $\Delta$. However, the SPH
interpolation does not produce the linear function exactly even on a uniformly
distributed set of particles unless $h/\Delta\to\infty$. However, if $h>\Delta$, 
the error becomes exponentially small. 

Now, the integral on the right side of Eq. (C.3) can be integrated by parts to
give
\begin{equation}
\int _{-\infty}^{\infty}\cos\left(2\pi j\frac{q}{\Delta}\right)W_{G}(q,h)dq=
-\frac{\Delta}{\pi j}\int _{0}^{\infty}\sin\left(2\pi j\frac{q}{\Delta}\right)
\frac{\partial W_{G}}{\partial q}dq,
\end{equation}
which can be bounded as follows
\begin{eqnarray}
\left|\int _{-\infty}^{\infty}\cos\left(2\pi j\frac{q}{\Delta}\right)W_{G}(q,h)dq\right|
&\leq&\frac{\Delta}{\pi j}\left|\int _{0}^{\infty}\sin\left(2\pi j\frac{q}{\Delta}\right)
\frac{\partial W_{G}}{\partial q}dq\right|\nonumber\\
&\leq&\frac{\Delta}{\pi j}\left|\sin\left(2\pi j\frac{q}{\Delta}\right)\right|
\left|\int _{0}^{\infty}\frac{\partial W_{G}}{\partial q}dq\right|\nonumber\\
&\leq&\frac{\Delta}{\pi j}W_{G}(0,h)=\frac{\Delta}{\pi ^{3/2}hj},
\end{eqnarray}
after use of Eq. (C.1). Therefore, the error bound of the SPH interpolation (C.4) is
\begin{equation}
\sum _{j=1}^{\infty}\exp\left(-\frac{\pi ^{2}j^{2}h^{2}}{\Delta ^{2}}\right)
\leq\frac{\Delta}{\pi ^{3/2}h}\sum _{j=1}^{\infty}\frac{1}{j}.
\end{equation}
Since the Gaussian kernel has infinite support, it is customary to define ${\cal{N}}$
as the average number of particles within distance $\sqrt{2} h$, which is directly
related to the numerical resolution scale necessary for resolving sound waves in
the continuum limit $h\gg d_{n}$ of large neighbor numbers, where $d_{n}$ is the
nearest neighbor distance \cite{Dehnen12}. Then, it follows from Eq. (21) that
$\Delta =2\sqrt{2} h/{\cal{N}}$. Using this relation into the right side of 
inequality (C.7), regularizing the harmonic series in the sense of (B.11), and using 
Eq. (30) we obtain the asymptotic expansion for the error bound
\begin{equation}
\sum _{j=1}^{\infty}\exp\left(-\frac{\pi ^{2}j^{2}h^{2}}{\Delta ^{2}}\right)
\leq\frac{2\sqrt{2}}{\pi ^{3/2}}\lim _{{\cal{N}}\to\infty}\frac{1}{{\cal{N}}}
\sum _{j=1}^{{\cal{N}}}\frac{1}{j}\leq\frac{2\sqrt{2}}{\pi ^{3/2}}
\frac{(1+\gamma)}{{\cal{N}}}+O\left(\frac{1}{{\cal{N}}^{2}}\right),
\end{equation}
which converges to zero as ${\cal{N}}\to\infty$. This is consistent with the 
leftmost term converging exponentially to zero as $h/\Delta\to\infty$.

\setcounter{equation}{0}
\renewcommand\theequation{D.\arabic{equation}}
\section{Error bound for the kernel approximation in multidimensions}

An error bound for the kernel approximation in $n$-dimensions can be
derived following essentially the same steps described in Appendix A
for the one-dimensional case. The error made by the kernel approximation
is given by the first summation on the right side of Eq. (36), which for
convenience is written in the equivalent form
\begin{eqnarray}
{\cal{E}}^{(n)}_{KS}&=&\langle f({\bf x}_{a})\rangle -f({\bf x}_{a})
\nonumber\\
&+&\frac{1}{2}\int _{\Omega _{n}}\left[({\bf x}_{b}-{\bf x}_{a})\cdot\nabla
\right]^{2}f({\bf x}_{a})W(\parallel {\bf x}_{a}-{\bf x}_{b}\parallel ,h)
d^{n}{\bf x}_{b},
\end{eqnarray}
where only terms up to $l=2$ are retained.

To find a bound on the error define
\begin{equation}
[({\bf x}_{b}-{\bf x}_{a})\cdot\nabla]^{2}f(\xi)=[{\bf u}\cdot\nabla]^{2}f(\xi)\leq
\sum _{i=1}^{n}\left[u_{i}\frac{\partial}{\partial\xi}_{i}\right]^{2}f(\xi _{i})
\leq n^{2}k^{2}h^{2}|D^{2}f(\xi)|,
\end{equation}
where ${\bf\xi}\in\Omega _{n}$ and ${\bf u}={\bf x}_{b}-{\bf x}_{a}$. 
Moreover, defining
\begin{equation}
e_{r}^{(2,n)}=\frac{k^{2}n^{2}}{2}\sup _{{\bf\xi}\in\Omega _{n}}
\left|D^{2}f({\bf\xi})\right|\geq \frac{k^{2}n^{2}}{2}
\left|D^{2}f({\bf\xi})\right|,
\end{equation}
where the operator $D^{2}$ means any second-order derivative (pure or
mixed), the error bound follows as
\begin{eqnarray}
\left|{\cal{E}}^{(n)}_{KS}\right|&=&\frac{1}{2}\left|\int _{\Omega _{n}}
\left[{\bf x}_{ba}\cdot\nabla\right]^{2}f({\bf x}_{a})
W(\parallel {\bf x}_{ab}\parallel ,h)d^{n}{\bf x}_{b}\right|,
\nonumber\\
&\leq&\frac{1}{2}\int _{\Omega _{n}}\left|\left[{\bf x}_{ba}
\cdot\nabla\right]^{2}f({\bf x_{a}})\right|
W(\parallel {\bf x}_{ab}\parallel ,h)d^{n}{\bf x}_{b},\nonumber\\
&\leq&\frac{1}{2}n^{2}k^{2}h^{2}\left|D^{2}f({\bf\xi})\right|
\int _{\Omega _{n}}W(\parallel {\bf x}_{ab}\parallel ,h)
d^{n}{\bf x}_{b},\nonumber\\
&\leq&e_{r}^{(2,n)}h^{2},
\end{eqnarray}
where ${\bf x}_{ba}={\bf x}_{b}-{\bf x}_{a}$ and ${\bf x}_{ab}=-{\bf x}_{ba}$.

\setcounter{equation}{0}
\renewcommand\theequation{E.\arabic{equation}}
\section{Error bound for the particle approximation in multidimensions}

As mentioned in the main text, the error when passing from the kernel approximation
to the particle approximation is quantified by the difference ${\cal{E}}_{KS}^{(n)}$
between the particle and the kernel estimates of a function $f({\bf x})$ evaluated at
the interpolation point ${\bf x}\in\mathbb{R}^{n}$. For a given particle $a$ at
${\bf x}_{a}$, this difference is given by the double summation in Eq. (36), which for
convenience is written in the equivalent form
\begin{equation}
{\cal{E}}_{KS}^{(n)}=
\sum _{\substack{{\bf j}\in\Lambda ^{\star}\\{\bf j}\neq {\bf 0}}}^{\infty}
\sum _{l=0}^{\infty}\frac{1}{l!}
\int _{\Omega _{n}}\left[({\bf x}_{b}-{\bf x}_{a})\cdot\nabla\right]^{l}f({\bf x}_{a})
W_{ab}\exp (-i2\pi {\bf j}\cdot {\bf b})
d^{n}{\bf x}_{b},
\end{equation}
where $W_{ab}=W(\parallel {\bf x}_{a}-{\bf x}_{b}\parallel ,h)$. To obtain a bound on this
error define
\begin{equation}
\left[({\bf x}_{b}-{\bf x}_{a})\cdot\nabla\right]^{l}f(\xi)=
\left[{\bf u}\cdot\nabla\right]^{l}f(\xi)\leq
\sum _{i}^{n}\left[\left|u_{i}\frac{\partial}{\partial\xi _{i}}\right|\right]^{l}f(\xi _{i})
\leq n^{l}k^{l}h^{l}|D^{l}f({\bf\xi})|,
\end{equation}
where ${\bf\xi}\in\Omega _{n}$, ${\bf u}={\bf x}_{b}-{\bf x}_{a}$, and $D^{l}$ denotes
any $l$th-order pure or mixed derivative. Now defining
\begin{equation}
{\tilde e}_{r}^{(l,n)}=\frac{k^{l}n^{l}}{l!}\sup _{{\bf\xi}\in\Omega _{n}}|D^{l}f({\bf\xi})|
\geq\frac{k^{l}n^{l}}{l!}|D^{l}f({\bf\xi})|,
\end{equation}
and the bound of the kernel function as
\begin{equation}
W(\parallel {\bf x}_{a}-{\bf x}_{b}\parallel ,h)\leq\frac{a_{0}}{h^{n}},
\end{equation}
where $a_{0}>0$, the bound on ${\cal{E}}_{KS}^{(n)}$ can be calculated as follows
\begin{eqnarray}
\left|{\cal{E}}_{KS}^{(n)}\right|&=&
\left|\sum _{\substack{{\bf j}\in\Lambda ^{\star}\\{\bf j}\neq {\bf 0}}}^{\infty}
\sum _{l=0}^{\infty}\frac{1}{l!}
\int _{\Omega _{n}}\left[{\bf x}_{ba}\cdot\nabla\right]^{l}f({\bf x}_{a})
W_{ab}\exp (-i2\pi {\bf j}\cdot {\bf b})d^{n}{\bf x}_{b}\right|\nonumber\\
&\leq&\sum _{\substack{{\bf j}\in\Lambda ^{\star}\\{\bf j}\neq {\bf 0}}}^{\infty}
\sum _{l=0}^{\infty}\frac{1}{l!}
\left|\int _{\Omega _{n}}\left[{\bf x}_{ba}\cdot\nabla\right]^{l}f({\bf x}_{a})
W_{ab}\exp (-i2\pi {\bf j}\cdot {\bf b})d^{n}{\bf x}_{b}\right|,
\end{eqnarray}
where $W_{ab}=W(\parallel {\bf x}_{ab}\parallel ,h)$, ${\bf x}_{ab}={\bf x}_{a}-{\bf x}_{b}$,
and ${\bf x}_{ab}=-{\bf x}_{ba}$. Let ${\bf x}_{b,\max}\in\Omega _{n}$ define a point where
$[{\bf x}_{b,\max,a}\cdot\nabla]^{l}f({\bf x}_{a})W(\parallel {\bf x}_{ab,\max}\parallel ,h)
=\max\{[{\bf x}_{ba}\cdot\nabla]^{l}f({\bf x}_{a})W(\parallel {\bf x}_{ab}\parallel ,h)\}$
and ${\bf x}_{ab,\max}={\bf x}_{a}-{\bf x}_{b,\max}$. Applying the Cauchy-Schwarz inequality
to (E.5) yields
\begin{eqnarray}
\left|{\cal{E}}_{KS}^{(n)}\right|&\leq&
\sum _{\substack{{\bf j}\in\Lambda ^{\star}\\{\bf j}\neq {\bf 0}}}^{\infty}
\sum _{l=0}^{\infty}\frac{1}{l!}
\left|\left[{\bf x}_{b,\max,a}\cdot\nabla\right]^{l}f({\bf x}_{a})\right|W_{ab,\max}
\left|\int _{\Omega _{n}}\exp (-i2\pi {\bf j}\cdot {\bf b})d^{n}{\bf x}_{b}\right|\nonumber\\
&\leq&a_{0}\sum _{l=0}^{\infty}h^{l-n}{\tilde e}_{r}^{(l,n)}
\sum _{\substack{{\bf j}\in\Lambda ^{\star}\\{\bf j}\neq {\bf 0}}}^{\infty}
\left|\int _{\Omega _{n}}\exp (-i2\pi {\bf j}\cdot {\bf b})d^{n}{\bf x}_{b}\right|,
\end{eqnarray}
where $W_{ab,\max}=W(\parallel {\bf x}_{ab,\max}\parallel ,h)$ and relation (E.3) has been 
used together with the definitions $|{\bf x}_{b,\max,a}|\leq k^{l}h^{l}$ and
$W(\parallel {\bf x}_{ab}\parallel ,h)\leq a_{0}/h^{n}$.

A bound to the Fourier integral when ${\cal{N}}\gg 1$ can be found by expanding
the vector function ${\bf b}({\bf x}_{b})$ about ${\bf x}_{a}$ to yield
\begin{equation}
{\bf b}({\bf x}_{b})={\bf b}({\bf x}_{a})+{\bf J}_{{\bf x}_{b}}({\bf x}_{a})\cdot
({\bf x}_{b}-{\bf x}_{a})+O(\parallel {\bf x}_{b}-{\bf x}_{a}\parallel ^{2})
\end{equation}
for ${\bf x}_{b}$ close to ${\bf x}_{a}$, where $\parallel {\bf x}_{b}-{\bf x}_{a}\parallel$
is the distance between ${\bf x}_{b}$ and ${\bf x}_{a}$ in the $n$-dimensional 
Euclidean space. Since this expression defines a linear mapping 
$\mathbb{R}^{n}\to\mathbb{R}^{m}$ with $m=n$, the determinant of the Jacobian matrix
${\bf J}_{{\bf x}_{b}}({\bf x}_{a})$ is given by relation (35), which from use of 
Eqs. (39) and (41) can be written as 
\begin{equation}
\left|{\bf J}_{{\bf x}_{b}}({\bf x}_{a})\right|=\frac{\rho ({\bf x}_{a})}
{m({\bf x}_{a},h)}=\frac{n{\cal{N}}}{B_{n}k^{n}h^{n}}=\frac{{\cal{N}}}{{\cal{V}}_{n}}=
\frac{1}{\Delta _{m}^{n}},
\end{equation}
for $kh\ll 1$, where ${\cal{V}}_{n}=B_{n}k^{n}h^{n}/n$ is the $n$-dimensional volume
of the kernel support and $\Delta _{m}$ is the mean distance of particle pairs within
the kernel support. Note that for low-discrepancy sequences of sample points 
$\Delta _{m}\approx\Delta _{s}$, where $\Delta _{s}$ (for $s=1,2,\dots ,n$) is the
projected mean distance on the $s$th-axis of an $n$-dimensional Cartesian coordinate
system. Use of the above linear mapping then yields for ${\bf j}\cdot {\bf b}$ the 
expression
\begin{equation}
{\bf j}\cdot {\bf b}({\bf x}_{b})={\bf j}\cdot {\bf b}({\bf x}_{a})+
{\bf j}^{T}\cdot {\bf J}_{{\bf x}_{b}}({\bf x}_{a})\cdot ({\bf x}_{b}-{\bf x}_{a}),
\end{equation}
where ${\bf j}^{T}$ is a row vector and ${\bf x}_{b}-{\bf x}_{a}$ is a column vector.
Defining the vector ${\bf v}$ as
\begin{equation}
{\bf v}={\bf J}_{{\bf x}_{b}}({\bf x}_{a})\cdot ({\bf x}_{b}-{\bf x}_{a}),
\end{equation}
and noting that $d^{n}{\bf v}=[n{\cal{N}}/(B_{n}k^{n}h^{n})]d^{n}{\bf x}_{b}$ by
Eq. (E.8), the Fourier integral in inequality (E.6) becomes
\begin{eqnarray}
\int _{\Omega _{n}}\exp (-i2\pi {\bf j}\cdot {\bf b})d^{n}{\bf x}_{b}&=&
\exp [-i2\pi {\bf j}\cdot {\bf b}({\bf x}_{a})]\frac{B_{n}k^{n}h^{n}}{n{\cal{N}}}
\nonumber\\
&\times&\int _{\tilde\Omega _{n}}\exp (-i2\pi {\bf j}\cdot {\bf v})d^{n}{\bf v},
\end{eqnarray}
where $\tilde\Omega =\tilde\Omega ({\bf v},h)$ is the image domain of
$\Omega =\Omega ({\bf x}_{b},h)$ due to the mapping ${\bf x}_{b}\to {\bf v}$.
If the Fourier exponential oscillates rapidly within the domain $\Omega _{n}$,
then the Fourier integral in (E.11) vanishes. For ${\cal{N}}\gg 1$, this
condition allows to test its convergence to zero. Now, expressing ${\bf j}\cdot {\bf b}$ 
in component form and noting that $d^{n}{\bf v}=dv_{1}dv_{2}\cdots dv_{n}$, the integral on 
the right side of (E.11) can be rewritten as
\begin{eqnarray}
\int _{\tilde\Omega _{n}}\exp (-i2\pi {\bf j}\cdot {\bf v})d^{n}{\bf v}&=&
\int _{\tilde\Omega _{n}}\exp\left(-i2\pi\sum_{s=1}^{n}j_{s}v_{s}\right)d^{n}{\bf v}
\nonumber\\
&=&\prod _{s=1}^{n}\int _{-c}^{c}\exp (-i2\pi j_{s}v_{s})dv_{s},
\end{eqnarray}
where $j_{s}$ and $v_{s}$ are, respectively, the projections of vectors ${\bf j}$ and
${\bf v}$ on the $s$th-axis of an $n$-dimensional Cartesian coordinate system. The
limits of integration in the second equality of Eq. (E.12) can be estimated as
follows. First, calculate a bound on the magnitude of vector ${\bf v}$ from Eq. (E.10)
as
\begin{equation}
v\leq kh\parallel {\bf J}_{{\bf x}_{b}}({\bf x}_{a})\parallel _{\infty}=
kh\max _{i}\sum _{j=1}^{n}|J_{ij}|=\frac{kh}{\Delta _{m}}\approx\frac{kh}{\Delta _{s}},
\end{equation}
where $\parallel {\bf J}_{{\bf x}_{b}}({\bf x}_{a})\parallel _{\infty}$ is the max
norm of the Jacobian matrix defined as the maximum absolute row sum of its elements 
$J_{ij}$. Note that the last equality in Eq. (E.13) holds because according to 
Eq. (E.8) the Jacobian is a diagonal matrix with elements $J_{ii}=1/\Delta _{m}$.
Since $\Delta _{m}\approx\Delta _{s}$ for a low-discrepancy sequence of sample points,
it follows that the projected components of vector ${\bf v}$ on the $s$th-axis of
the $n$-dimensional Cartesian coordinate system satisfy the bound inequality
$|v_{s}|\leq c=kh_{s}/\Delta _{s}$, where $h_{s}$ is the projection of $h$ on the
$s$th-axis of the Cartesian system. Following similar steps to those described 
in Appendix B for the one-dimensional case, it is easy to show that
\begin{equation}
\left|\int _{-c}^{c}\exp (-i2\pi j_{s}v_{s})dv_{s}\right|=
\left|\int _{-c}^{c}\cos (2\pi j_{s}v_{s})dv_{s}\right|\leq\frac{1}{\pi j_{s}},
\end{equation}
Since the maximum value of the cosine integral is $1/\pi j_{s}$, the bound on the Fourier 
integral takes the form
\begin{equation}
\sum _{\substack{{\bf j}\in\Lambda ^{\star}\\{\bf j}\neq {\bf 0}}}^{\infty}
\left|\int _{\Omega _{n}}\exp (-i2\pi {\bf j}\cdot {\bf b})d^{n}{\bf x}_{b}\right|
\leq\left(\frac{2}{\pi}\right)^{n}\frac{B_{n}k^{n}h^{n}}{n}\left(\lim _{{\cal{N}}\to\infty}
\frac{1}{{\cal{N}}}\prod _{s=1}^{n}\sum _{j_{s}=1}^{{\cal{N}}_{s}}
\frac{1}{j_{s}}\right),
\end{equation}
where ${\cal{N}}_{s}=[2kh_{s}/\Delta _{s}]$. With this result, the bound on 
${\cal{E}}_{KS}^{(n)}$ can be finally written as
\begin{equation}
\left|{\cal{E}}_{KS}^{(n)}\right|\leq\left(\frac{2}{\pi}\right)^{n}
\frac{a_{0}B_{n}k^{n}}{n}\sum _{l=0}^{\infty}h^{l}{\tilde e}_{r}^{(l,n)}
\left(\lim _{{\cal{N}}\to\infty}
\frac{1}{{\cal{N}}}\prod _{s=1}^{n}\sum _{j_{s}=1}^{{\cal{N}}_{s}}\frac{1}{j_{s}}\right),
\end{equation}
which demonstrates inequality (43). Noting that $\Delta _{s}\approx\Delta _{\rm m}$ for
a low-discrepancy set of sample points, it follows from relation (41) that
${\cal{N}}_{s}\propto {\cal{N}}^{1/n}$ and hence the limit between parentheses in 
inequality (E.16) converges in the sense of (B.11).
\end{appendices}

\section*{Acknowledgement}
We are grateful to the anonymous reviewers for their valuable comments and suggestions.
This work was partially supported by the Conacyt ENERXICO project under the grant
number B-S-6992, the Division of Basic Sciences and Engineering (CBI) of the 
Autonomous Metropolitan University (UAM-A) and the Venezuelan Institute of Scientific
Research (IVIC) through internal funds. The calculations of this paper were performed 
using the computing facilities of Cinvestav-Abacus.

\section*{References}

\bibliography{mybibfile}

\clearpage

\begin{figure}
\caption{Dependence of the 2D (solid circles) and 3D (solid squares) error
bounds given by relations (56) and (57), respectively, and the mean
absolute errors (MAEs) between the SPH estimates of the test functions (53)
(open circles) and (54) (open squares) and their exact values on
spatial resolution $N^{1/n}$, where $n=2$ in 2D and $n=3$ in 3D.}
\end{figure}

\begin{figure}
\caption{Distributions of the volume estimate $M_{0}$ for the 2D quasi-random
point set for the spatial resolutions listed in Table 1. The distributions of
$M_{0}$ slowly approach a narrow normal distribution peaking at $M_{0}=1$
with increasing $N$ and $\mathcal{N}$. The left inset depicts the maximum
of the distributions as a function of $N^{1/2}$ and the right inset shows
the details of the distributions for $N\leq 100^{2}$. This quantifies the
deviation from an exact partition of unity.}
\end{figure}

\begin{figure}
\caption{Distributions of the volume estimate $M_{0}$ for the 3D quasi-random
point set for the spatial resolutions listed in Table 1. The distributions of
$M_{0}$ slowly approach a narrow normal distribution peaking at $M_{0}=1$
with increasing $N$ and $\mathcal{N}$. The left inset depicts the maximum of
the distributions as a function of $N^{1/3}$ and the right inset shows the
details of the distributions for $N\leq 100^{3}$. This quantifies the
deviation from an exact partition of unity.}
\end{figure}

\clearpage

\begin{figure}
\includegraphics[angle=-90,width=12cm]{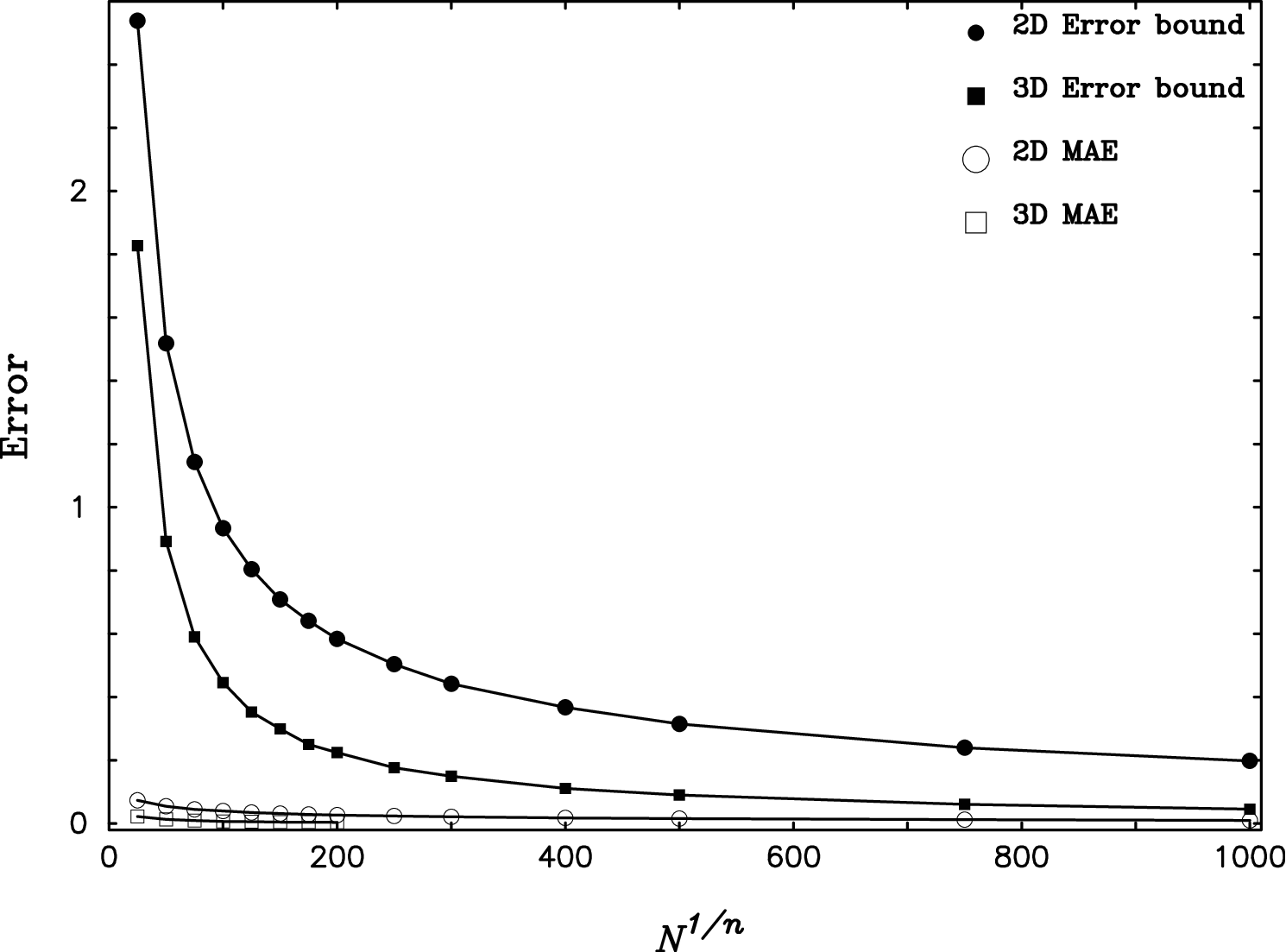}
\end{figure}
Figure 1

\clearpage

\begin{figure}
\includegraphics[width=12cm]{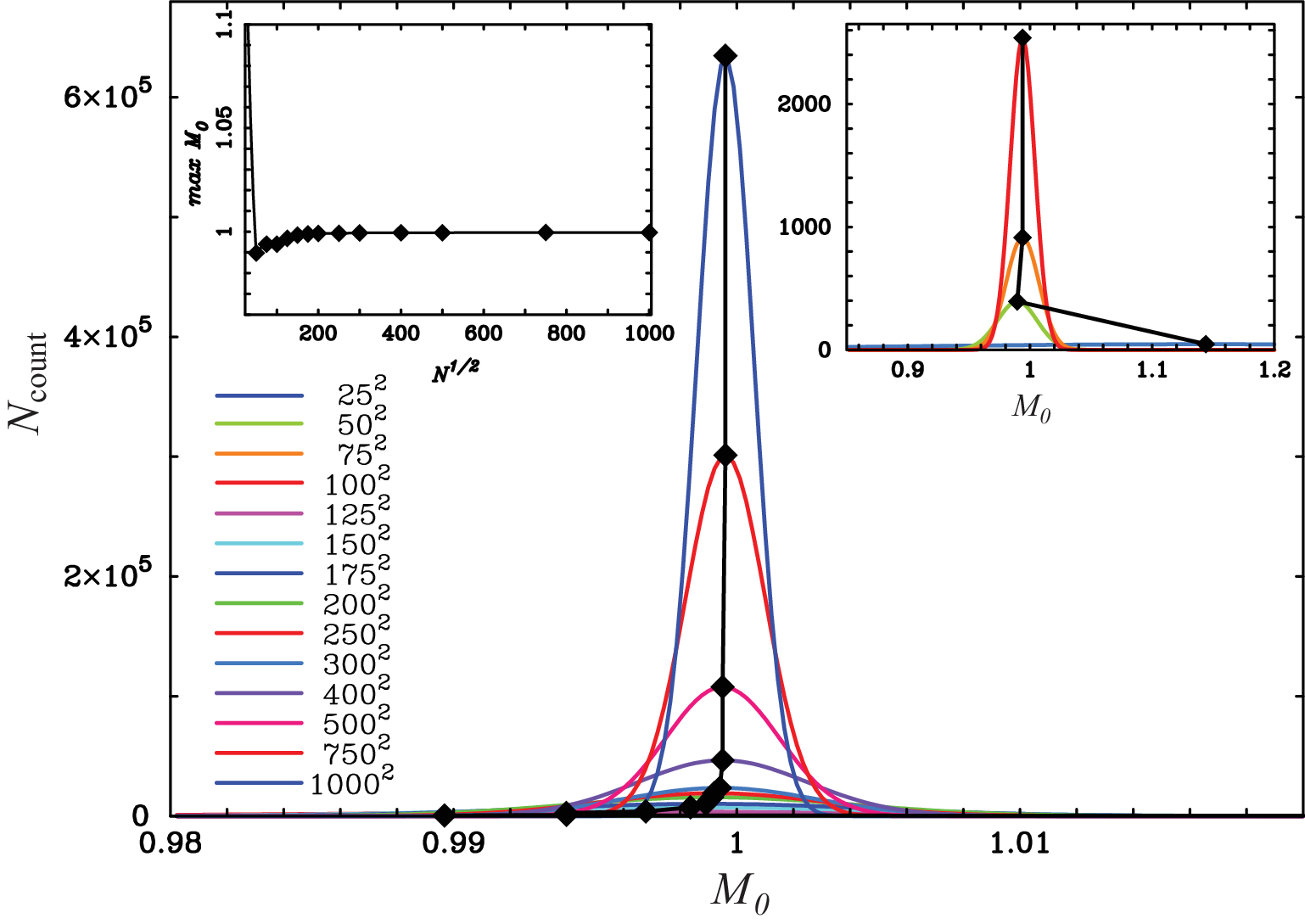}
\end{figure}
Figure 2

\clearpage

\begin{figure}
\includegraphics[width=12cm]{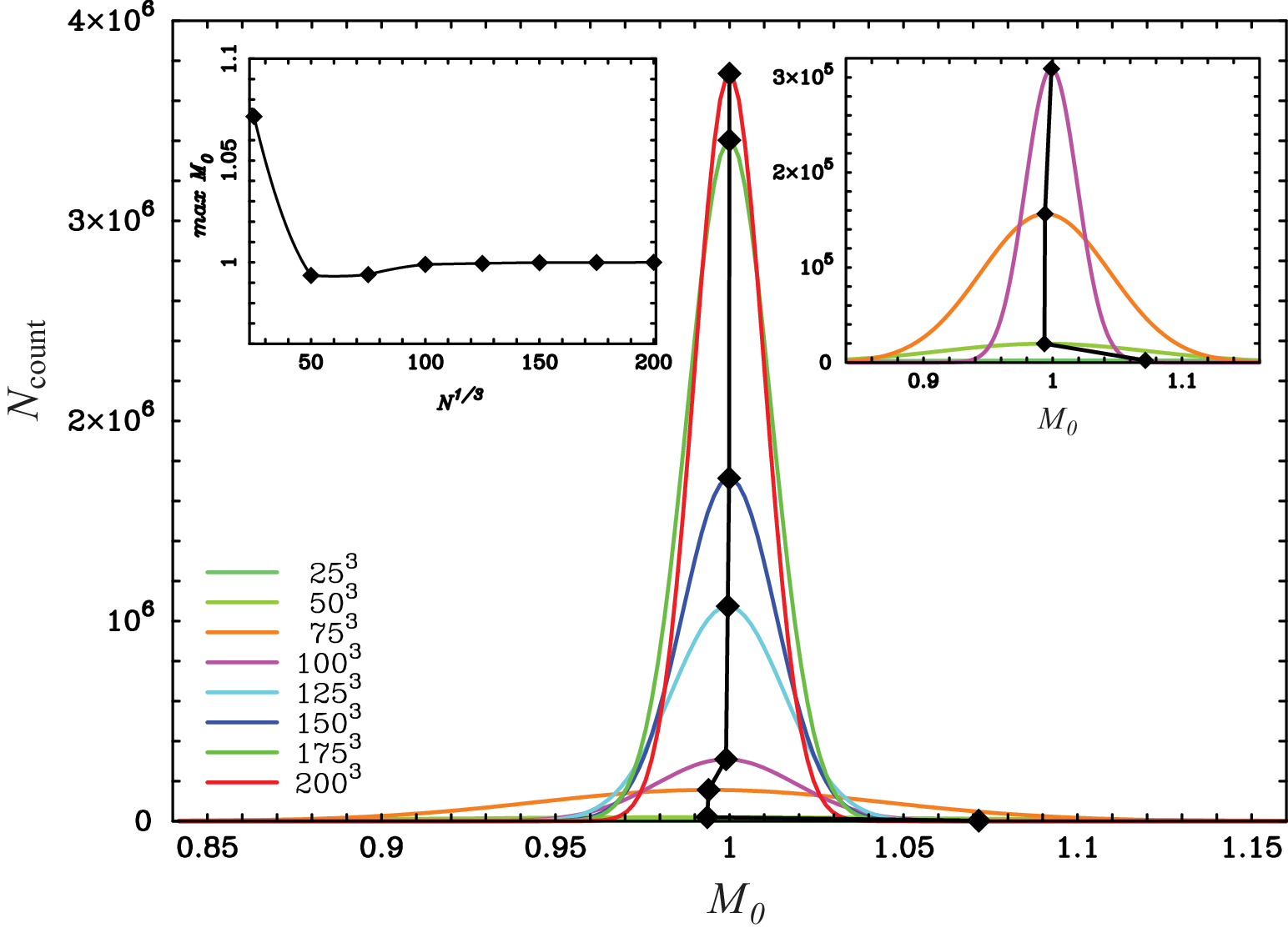}
\end{figure}
Figure 3

\end{document}